\documentclass[article, nojss]{jss}

\author{Mark O'Connell\\Maynooth University \And
        Catherine B. Hurley\\Maynooth University \And
        Katarina Domijan\\Maynooth University}
\title{Conditional Visualization for Statistical Models:
  \\An Introduction to the \pkg{condvis} Package in \proglang{R}}

\Plainauthor{Mark O'Connell, Catherine B. Hurley, Katarina Domijan}
\Plaintitle{Conditional Visualization for Statistical Models:
  An Introduction to the condvis Package in R}
\Shorttitle{\pkg{condvis}: Conditional Visualization for Statistical Models}

\Abstract{
  The \pkg{condvis} package is for interactive visualization of sections in data
  space, showing fitted models on the section, and observed data near the
  section. The primary goal is the interpretation of complex models, and showing
  how the observed data support the fitted model. There is a video accompaniment
  to this paper available  \href{https://www.youtube.com/watch?v=rKFq7xwgdX0}
  {here}. This is a preprint version of \citet{myjsspaper}.
}
\Plainkeywords{interactive, graphics, regression, classification,
  blackbox models}

\Address{
  Mark O'Connell\\
  Department of  Mathematics \& Statistics\\
  Logic House\\
  Maynooth University\\
  Co. Kildare, Ireland\\
  E-mail: \email{mark.oconnell@maths.nuim.ie}\\
}

\usepackage{lmodern}
\usepackage{amsmath}
\usepackage{amssymb}
\usepackage{natbib}
\usepackage{float}
\usepackage{subcaption}

\newcommand{\predictors}{\boldsymbol{x}}

\newcommand{\lpnorm}[2]{\lvert\lvert #1 \rvert\rvert _{#2}}

\graphicspath{{Figures/}}

\begin{document}
\section{Introduction}
\begin{figure}
  \centering
  \includegraphics[width=0.3\textwidth]{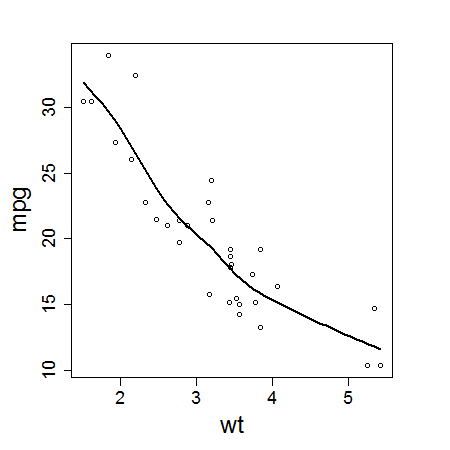}
  \includegraphics[width=0.3\textwidth]{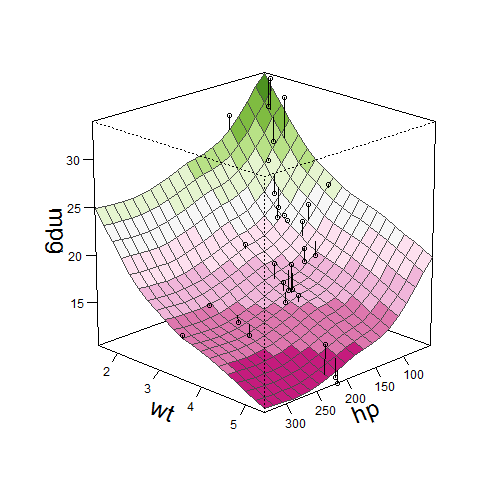}
  \hspace{5mm}
  \raisebox{5mm}{\includegraphics[width=0.25\textwidth]{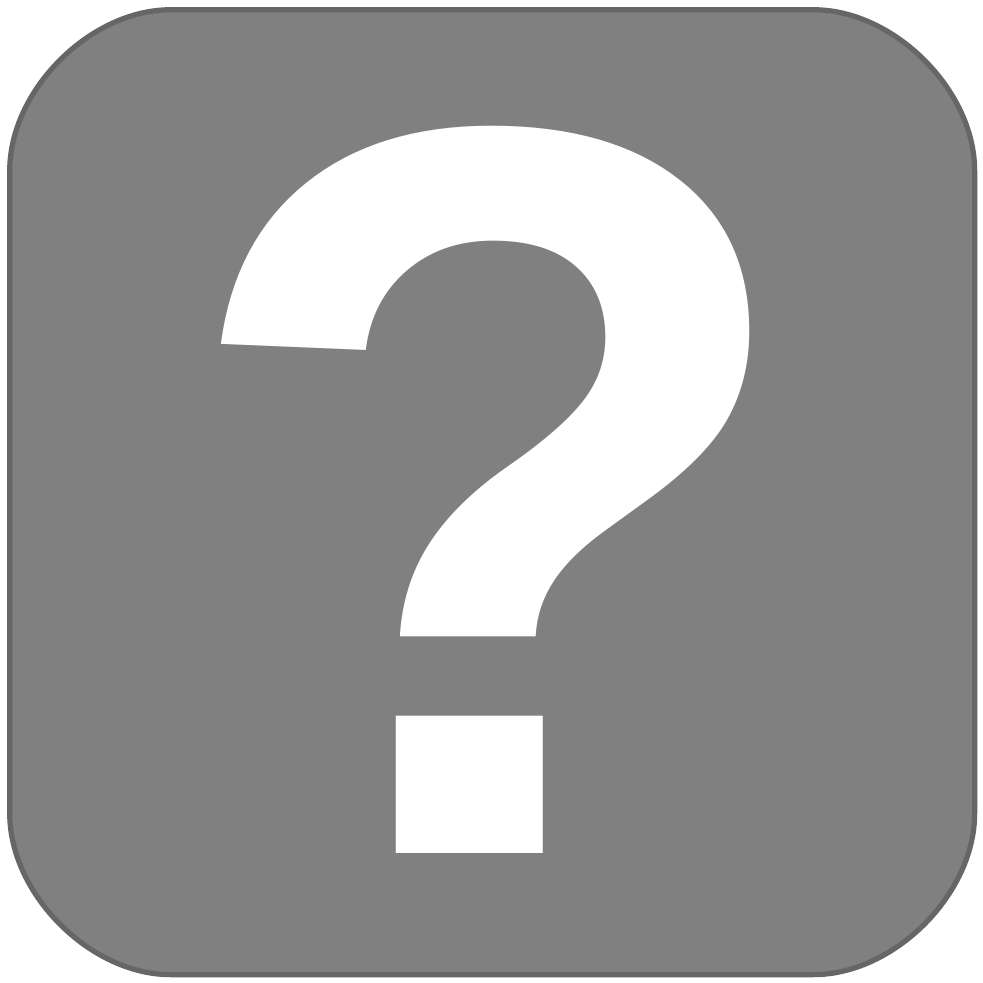}}
  \\\hspace{5mm} 1 predictor \hspace{25mm} 2 predictors \hspace{18mm} \ldots
    more predictors
  \caption{Visualizing fitted models. With one predictor, fitted model may be
  visualized as a curve. With two predictors, fitted model may be visualized as
  a surface in three dimensions. What can we do for more predictors?}
  \label{fig:modelvis}
\end{figure}

When a model consists of a single continuous predictor and a single response,
the fitted model is simply visualized as a curve in two dimensions.  When a
model involves two predictors, it may be visualized as a surface in three
dimensions; either as a contour plot or a perspective mesh. When a model
involves more than two predictors, there is no direct way to visualize the model
behaviour (see Figure~\ref{fig:modelvis}). Clearly, there is a need for
producing low-dimensional visualizations of models in high-dimensional space.
One approach is conditional visualization.
\par
\begin{figure}
  \centering
  \begin{subfigure}[b]{0.45\textwidth}
    \includegraphics[width = \textwidth]{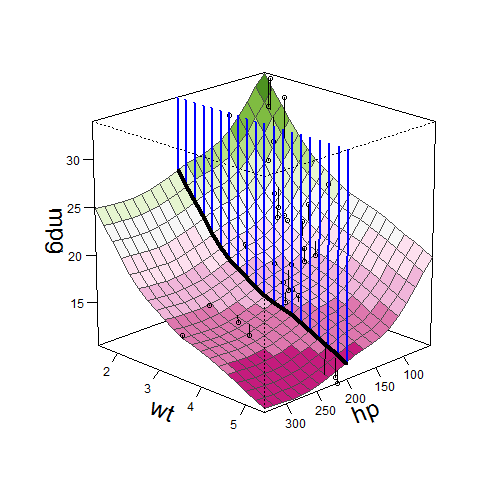}
    \caption{}
    \label{fig:section3d}
  \end{subfigure}
  \hspace{5mm}
  \begin{subfigure}[b]{0.45\textwidth}
    \includegraphics[width = \textwidth]{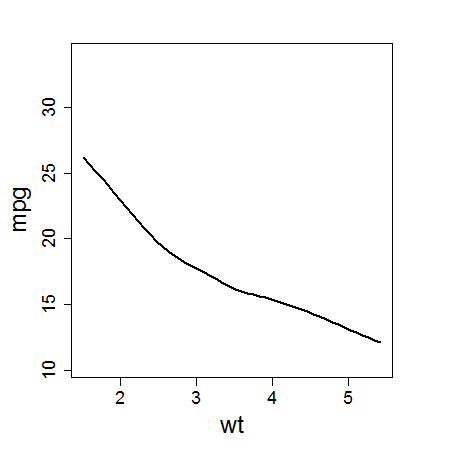}
    \caption{}
    \label{fig:section2d}
  \end{subfigure}
  \caption{Visualizing a section. (a) Visualizing a 2 predictor model, and
  showing a section at hp = 200. (b) Visualizing the section through model at hp
  = 200.}
  \label{fig:section}
\end{figure}

In a geometric sense, conditional visualization means taking a section. Consider
a simple model with two predictors, relating \code{mpg} to \code{wt} and
\code{hp} in the \code{mtcars} data in \proglang{R} \citep{R}. The fitted model
may be visualized as a surface as in Figure~\ref{fig:section3d}. If we want to
visualize the modelled effect of \code{wt} conditional on \code{hp}, we take a
section. The intersection of the fitted model and the section is then a curve in
two dimensions as in Figure~\ref{fig:section2d}. In this way, conditional
visualization offers a way to produce low-dimensional visualizations of models
in high-dimensional space. It is important to note that such sections typically
have no observed data lying on them, and so it is difficult to understand how
the observed data support the fitted model. In \pkg{condvis}, we choose to
display observed data points which are deemed to be \emph{near} the section
(discussed further in Section~\ref{sec:sections}).
\subsection{Example: Forced expiratory volume (FEV) data}
\label{sec:fev}
The FEV dataset in \citet{kahn05fev} provides some useful discussion material
for conditional relationships in statistical models. It originally appears in an
earlier edition of \citet{rosner2010fundamentals}; we use the copy provided in
the \pkg{covreg} \citep{covreg-package} package. We use this data to launch into
an example of the use of \pkg{condvis}. The dataset concerns the relationship
between lung health and smoking in children. The response is forced expiratory
volume (FEV, the amount of air an individual can exhale in the first second of
a forceful breath, used as a proxy for lung health), and the predictors are
gender, age, height and smoking status (binary).
\par
\begin{figure}
  \centering
  \includegraphics[width = 0.5\textwidth]{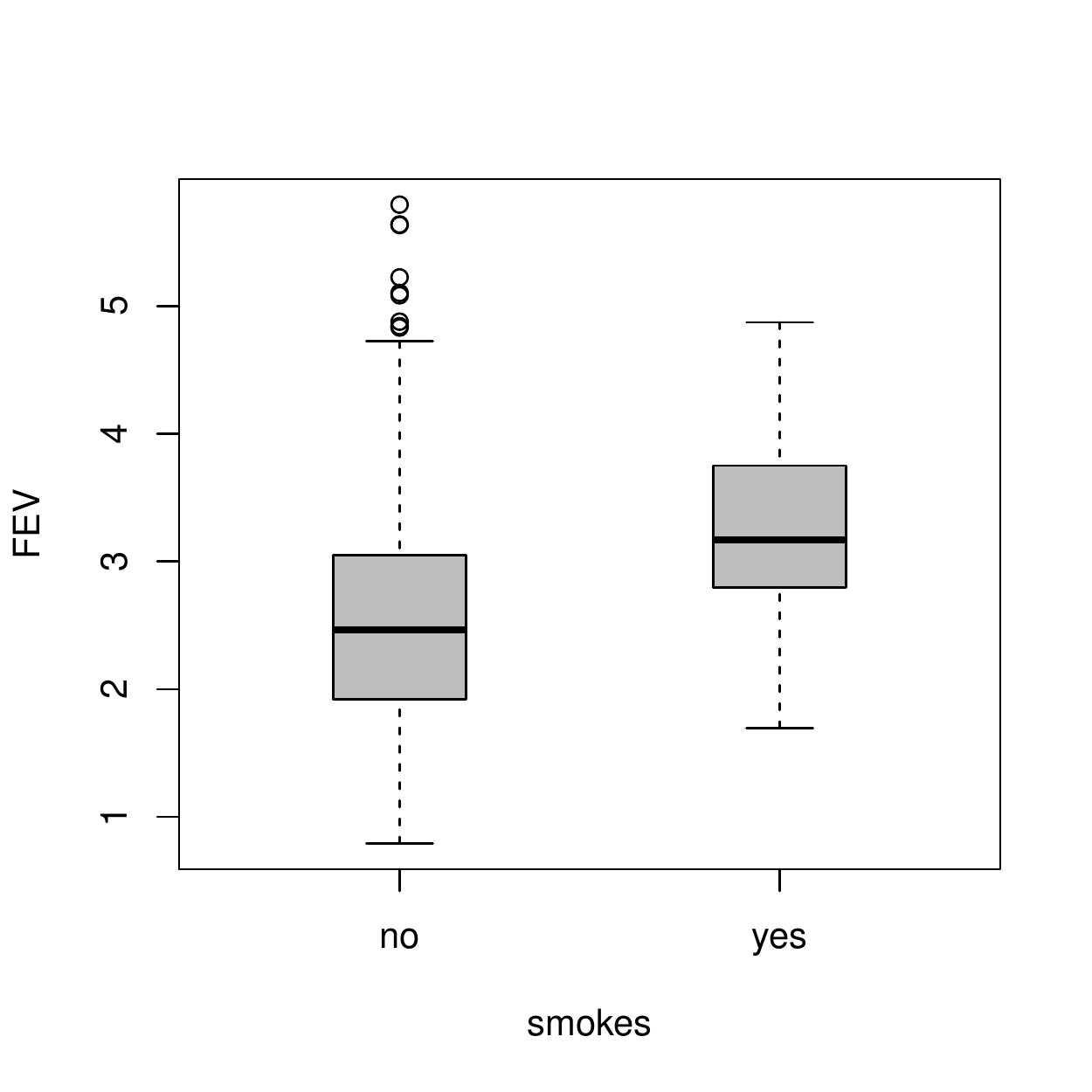}
  \caption{Boxplot of FEV versus smoking status. Suggests smoking is associated
  with higher FEV values.}
  \label{fig:ex1_box}
\end{figure}

In making a plot of FEV versus smoking status, we get our first surprise (see
Figure~\ref{fig:ex1_box}).
In the marginal view, it seems as though smoking is associated with better lung
health! To illustrate the use of \pkg{condvis}, we fit a support vector machine
\citep{smola1997support} from the \pkg{e1071} \citep{e1071-package} package.
\begin{Code}
R> library("e1071")
R> m1 <- svm(fev ~ gender + smoke + age + height, data = fev)
\end{Code}
We then call \code{ceplot} and start looking at sections through the fitted
model, investigating the modelled effect of smoking on FEV, conditional on the
other predictors. (There is a \href{https://youtu.be/rKFq7xwgdX0?t=270}{
video demonstration} of this example, and a Shiny application
\href{https://markajoc.github.io/condvis/example-fev.html}{demo}.)
\begin{Code}
R> ceplot(data = fev, model = m1, sectionvars = "smoke", type = "separate")
\end{Code}
\par
\begin{figure}
  \centering
  \includegraphics[width = 0.55\textwidth, clip = TRUE, trim = {0 0 0 15mm}]
    {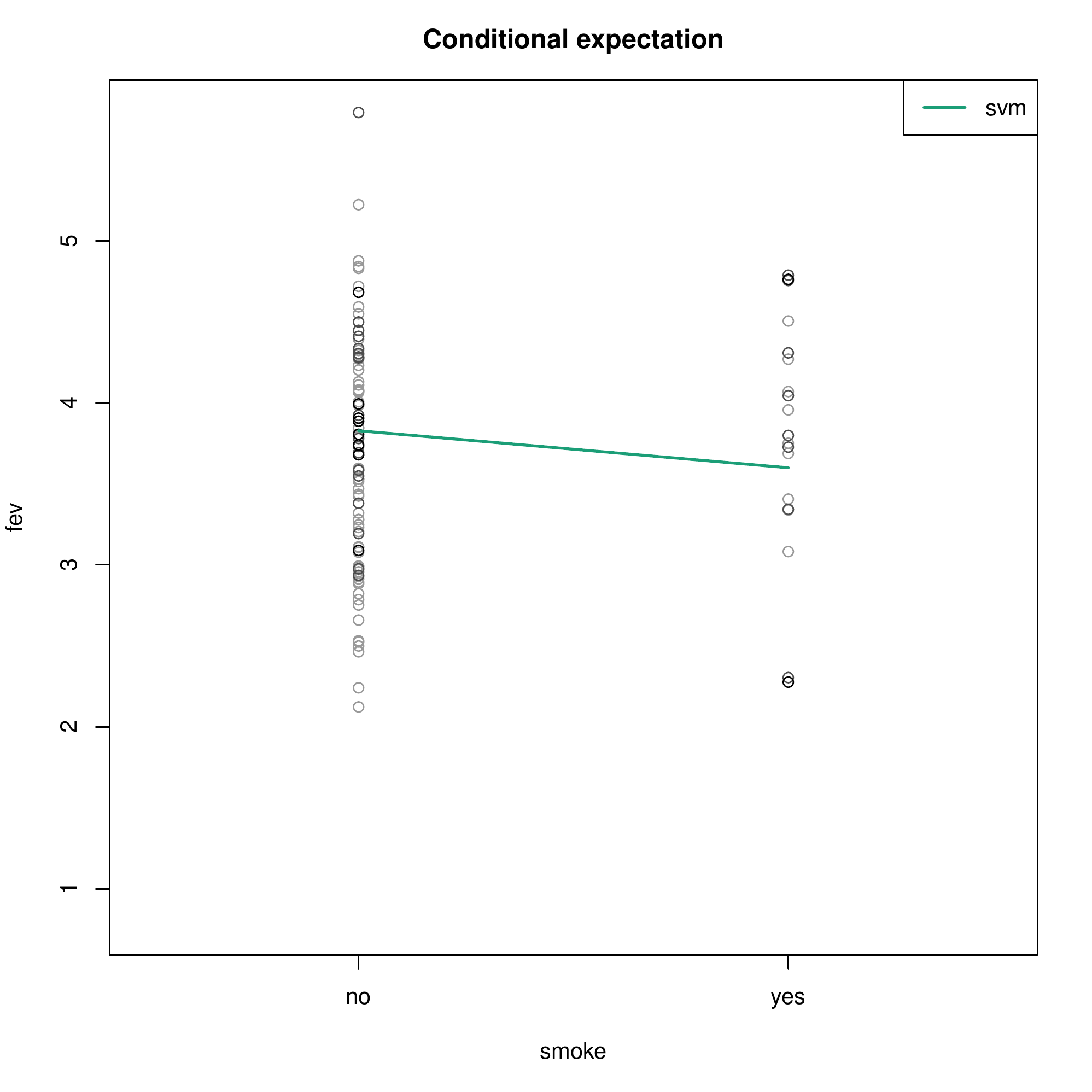}
  \raisebox{+20mm}{\includegraphics[width = 0.2\textwidth]
    {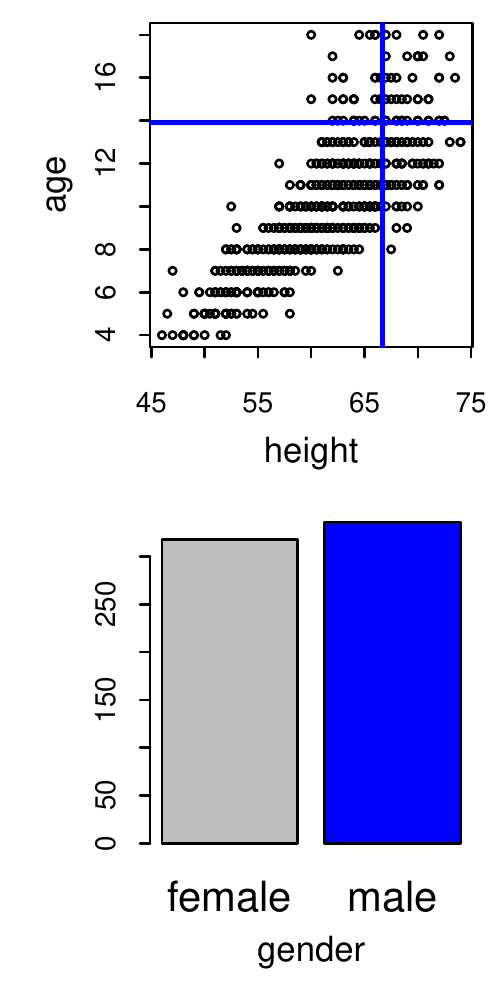}}
  \caption{Section showing the modelled effect of smoking on FEV conditional on
  height = 67, age = 14, gender = male. Suggests that smoking is associated with
  lower FEV values, and observed data near this section seem to support this.}
  \label{fig:ex1_sec1}
  \vspace{15mm}
  \includegraphics[width = 0.55\textwidth, clip = TRUE, trim = {0 0 0 15mm}]
    {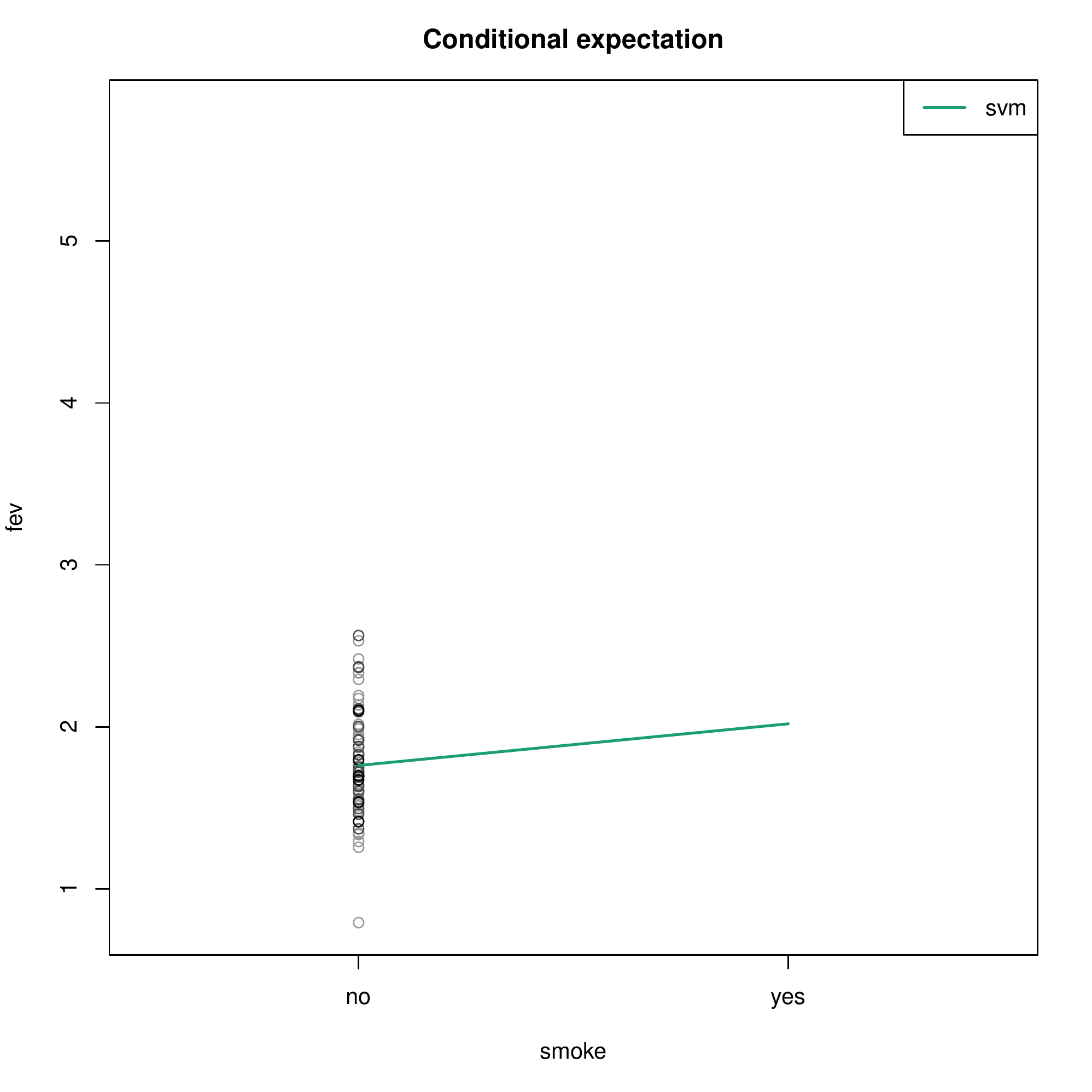}
  \raisebox{+20mm}{\includegraphics[width = 0.2\textwidth]
    {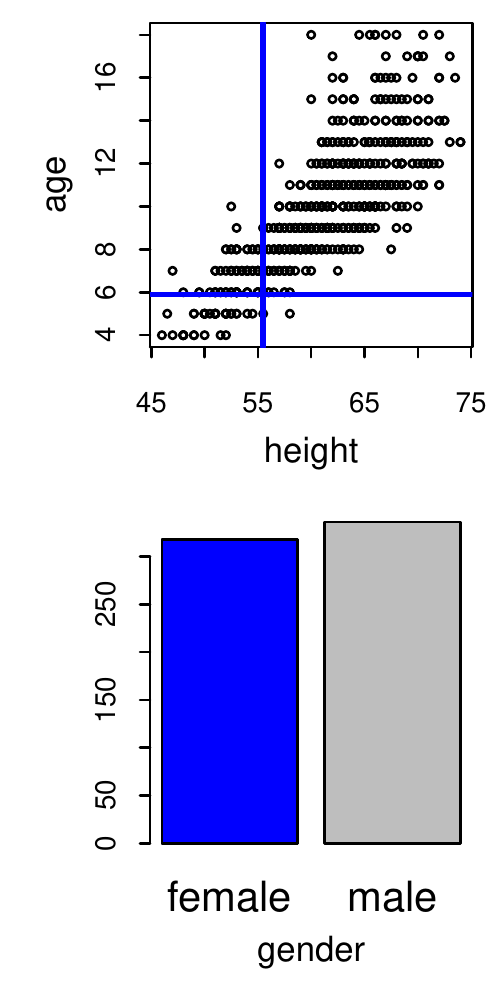}}
  \caption{Section showing the modelled effect of smoking on FEV conditional on
  height = 55, age = 6, gender = female. Model suggests that smoking is
  associated with higher FEV values, but there are no observations in the
  smoking group near this section.}
  \label{fig:ex1_sec2}
\end{figure}

Taking a section around age = 14, height = 67, with either gender, shows a more
sensible result (see Figure~\ref{fig:ex1_sec1}). In these parts of the predictor
space, the fitted model suggests that smoking is associated with slightly lower
FEV values. This is an example of Simpson's paradox, where the modelled
conditional association is of opposite sign to the apparent marginal
association. The observed data near these sections also seem to support the
fitted model, although it is worth noting that, for each section, there are
consistently more observations in the non-smoking group compared to the smoking
group.
\par
Taking a section around age = 6, height = 55, gender = female, the model is
suggesting that smoking is related to higher FEV values (see
Figure~\ref{fig:ex1_sec2}),
as in the marginal view before! Why is this? On examining the section, we see
there are no observed data points in the smoking group in this part of the data
space. It is not surprising that there are no 6 year old smokers. Such a
prediction for 6 year old smokers should clearly be considered an extrapolation
and held in suitable suspicion accordingly. This example demonstrates how
`black-box' models can mislead analysts just as easily as more rigid linear
models. A good defence against this is to take sections through the model and
see how nearby observed data support the model.
\subsection{Outline}
The remainder of the article proceeds as follows: Section~\ref{sec:overview}
gives an overview of the \pkg{condvis} package, explaining the basic ideas
behind it. Section~\ref{sec:using} gives more detail on how to use
\pkg{condvis}, describing available options and showing some code examples.
Section~\ref{sec:related} briefly discusses some other approaches to conditional
visualization in the literature and in \proglang{R}. Section~\ref{sec:summary}
concludes; giving a short summary, some strengths and limitations of
\pkg{condvis}, and finally an outlook on further work. The appendix provides
information on the supplementary materials.
\section{Overview}
\label{sec:overview}
%
\subsection{Basic workflow}
The \pkg{condvis} package is available from
\href{https://cran.r-project.org/web/packages/condvis/}{CRAN} and can be
installed using
\begin{Code}
R> install.packages("condvis")
\end{Code}
The source code is also hosted at \href{https://github.com/markajoc/condvis}
{Github}, where bug reports and feature requests are welcome.
\par
The main function in \pkg{condvis} is \code{ceplot}, which produces a
visualization consisting of two main parts; a section in data space, and
condition selector plots for interactively choosing the section. The default
behaviour is to place all graphics on a single device with base \proglang{R}
graphics, but options exist to place the section and condition selectors on
separate devices, and to produce a Shiny \citep{shiny} web application with the
same displays in a web browser.
\par
\begin{figure}[t]
  \centering
  \includegraphics[width = 0.8\textwidth]{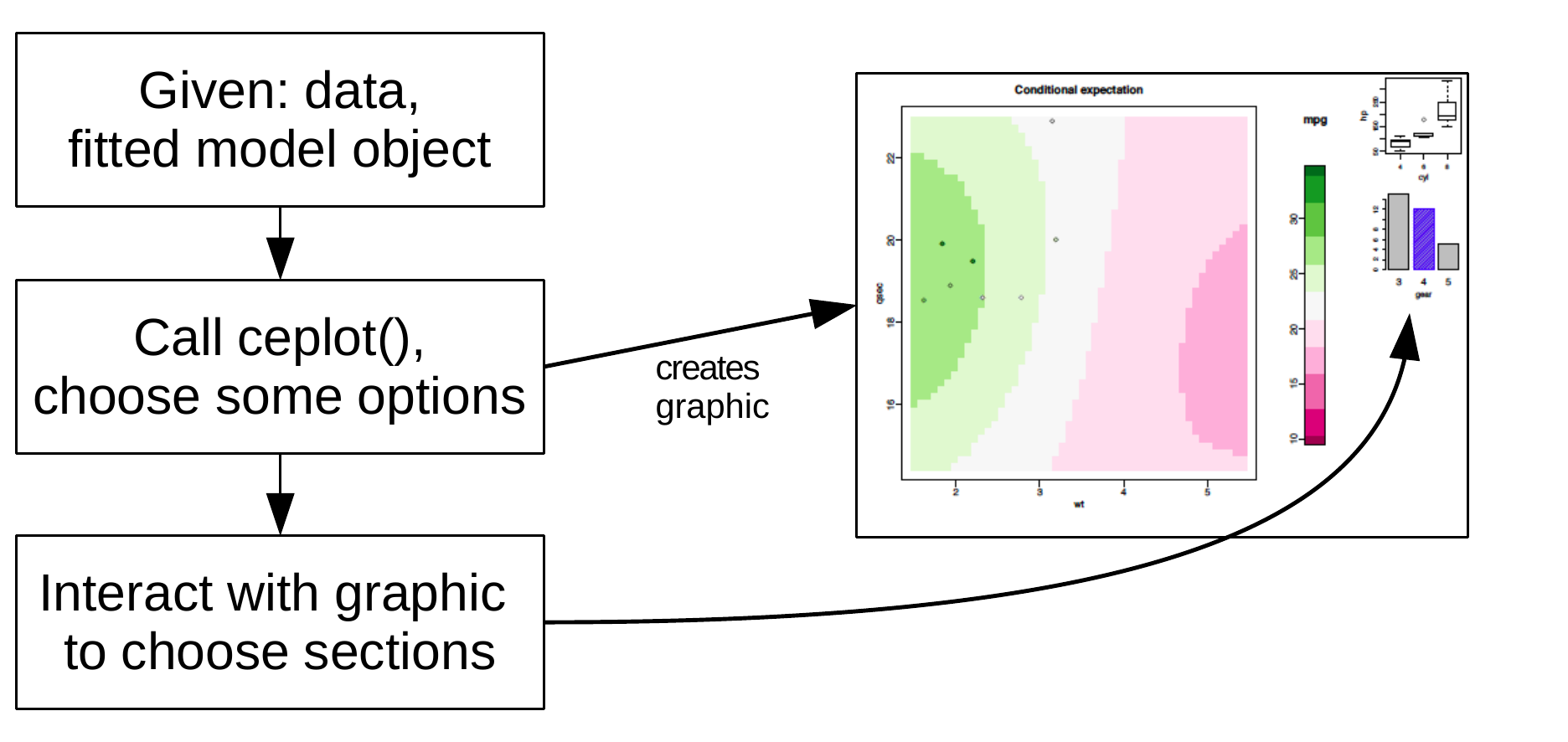}
  \caption{Basic workflow with \code{ceplot}.}
  \label{fig:workflow}
\end{figure}

Any usage of \code{ceplot} begins with a dataset and a fitted model object in
\proglang{R} (see Figure~\ref{fig:workflow}). This is minimally sufficient to
call \code{ceplot} if model terms can be extracted from the model object. In
most cases, we will want to specify the response and the predictor(s) of
interest, the predictor(s) along which we want to take a section. These are
specified as the character name of the variable in the dataframe.
\par
For a very simple example of the workflow in \pkg{condvis}, we can fit a linear
model to the \code{mtcars} data, and then call \code{ceplot}, specifying
\code{hp} as the predictor along which to take sections.
\begin{Code}
R> m2 <- lm(mpg ~ wt + hp, data = mtcars)
R> ceplot(data = mtcars, model = m2, sectionvars = "hp")
\end{Code}
In this case, the response variable can be extracted from the fitted model, and
so does not need to be specified.
\par
Sections through fitted models are evaluated using the generic \code{predict}
function, so this currently works for models with standard predict methods.
Models of class \code{lm}, \code{glm}, \code{gam}, \code{svm}
\citep{e1071-package}, and \code{randomForest} \citep{randomForest-package} are
examples of such models. For basis expansions in linear models, the \code{AsIs}
function (for example, \code{I(x\^{}2)}) must be used in the formula, unless the
user explicitly wants to investigate the model in the expanded data space. The
same goes for interactions, which should be specified using \code{a*b} or
\code{a:b} as necessary, rather than adding constructed variables to the
dataframe. If a fitted model object does not have a suitable \code{predict}
method, a S3 wrapper object may be used to standardize the model object's
external behaviour under a \code{predict} call.
\subsection{Choosing sections: Condition selector plots}
\label{sec:choosingsections}
In \pkg{condvis}, choosing a section is treated as choosing a point in the space
of the conditioning predictors. This could be accomplished by keyboard input or
sliders for continuous predictors, and a dropdown menu for categorical
predictors. Such a simple approach would quickly succumb to the curse of
dimensionality however, and it would be difficult to choose sensible sections.
By sensible, we mean that a section is roughly in a part of the data space where
there are observed data points, and not an unwitting extrapolation.
\par
\begin{figure}
  \centering
  \includegraphics[width = 0.3\textwidth]{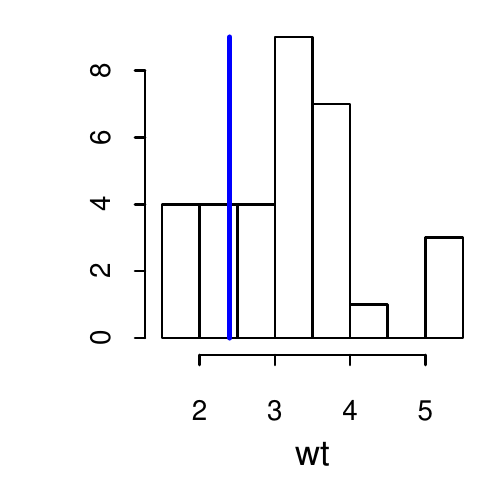}
  \includegraphics[width = 0.3\textwidth]{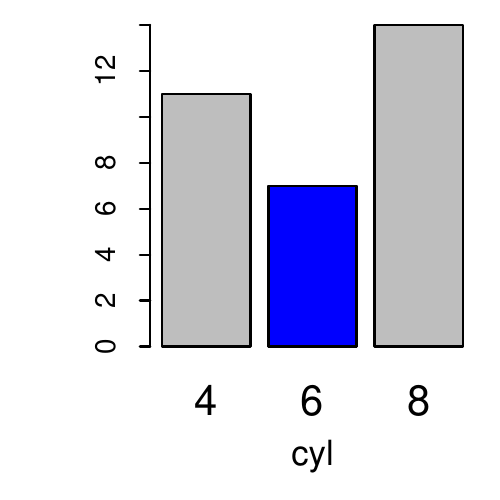}
  \caption{Univariate condition selector plots.}
  \label{fig:xcuniplots}
  \vspace{15mm}
  \includegraphics[width = 0.3\textwidth]{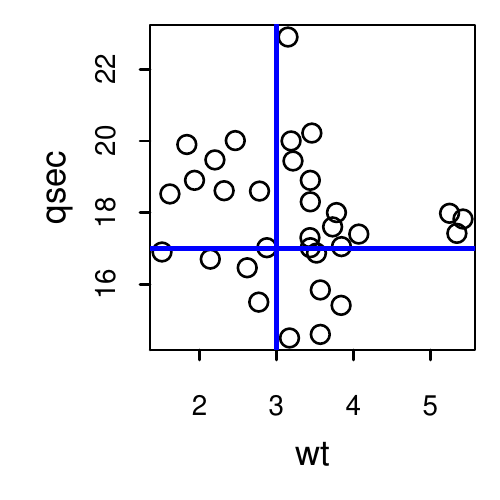}
  \includegraphics[width = 0.3\textwidth]{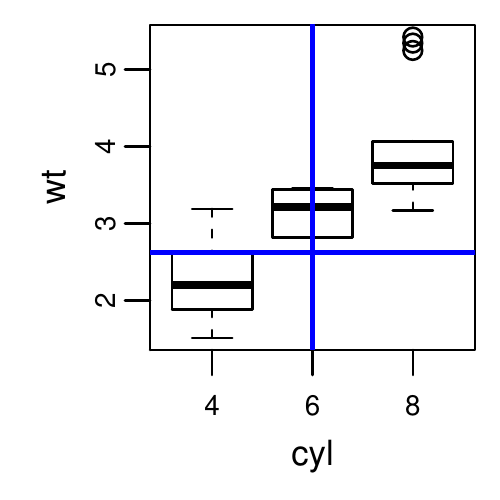}
  \includegraphics[width = 0.3\textwidth]{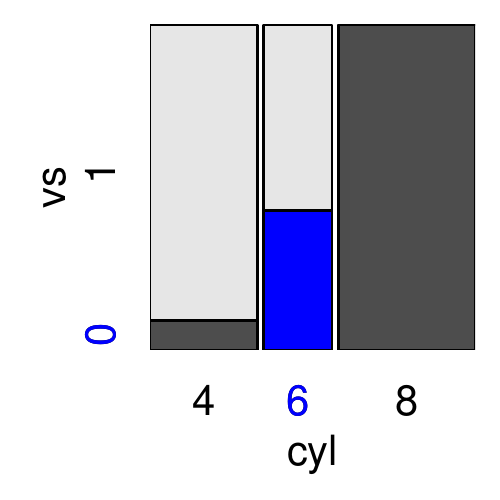}
  \caption{Bivariate condition selector plots.}
  \label{fig:xcbiplots}
  \vspace{10mm}
  \includegraphics[width = 0.45\textwidth]{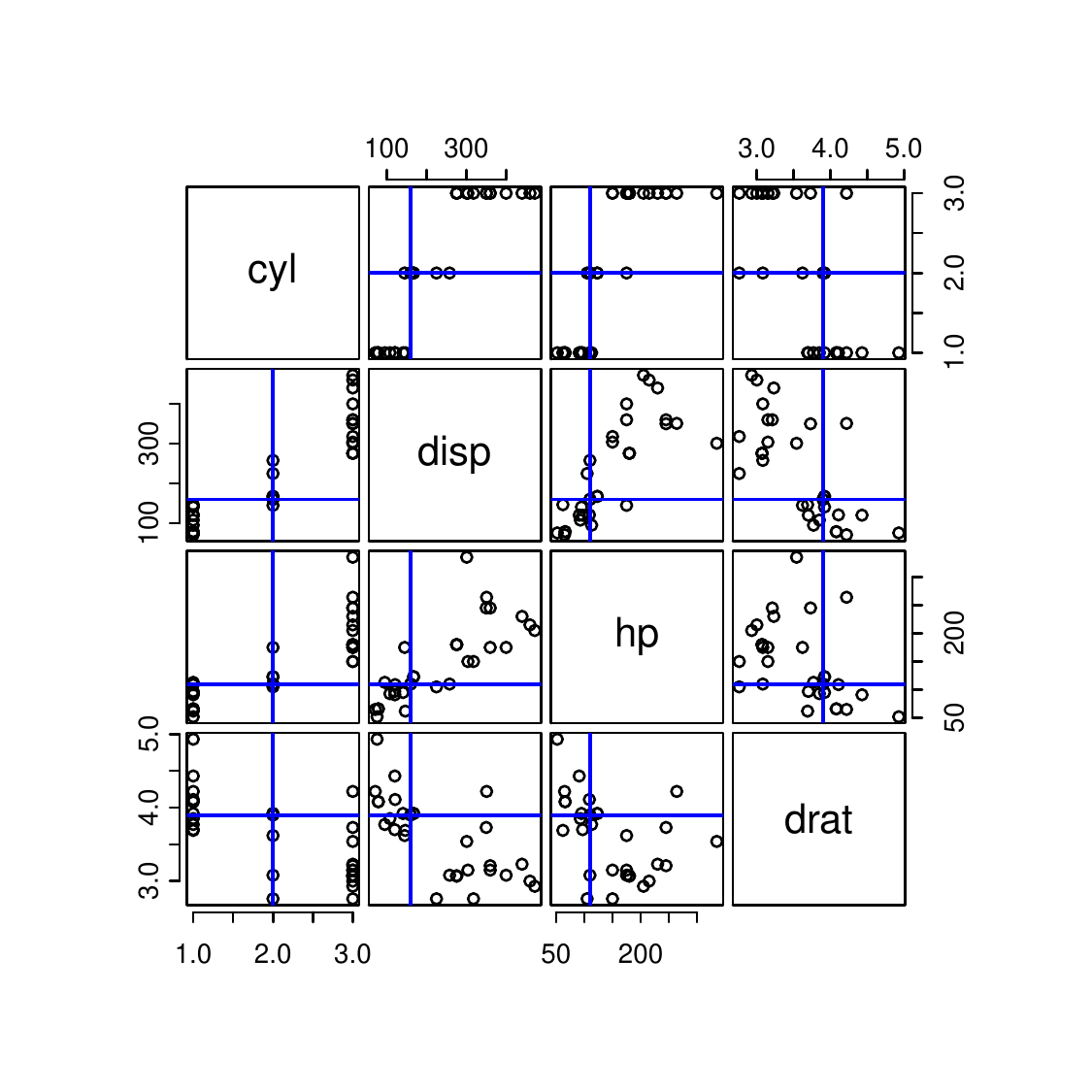}
  \raisebox{0.1\textwidth}{\includegraphics[width = 0.45\textwidth]
    {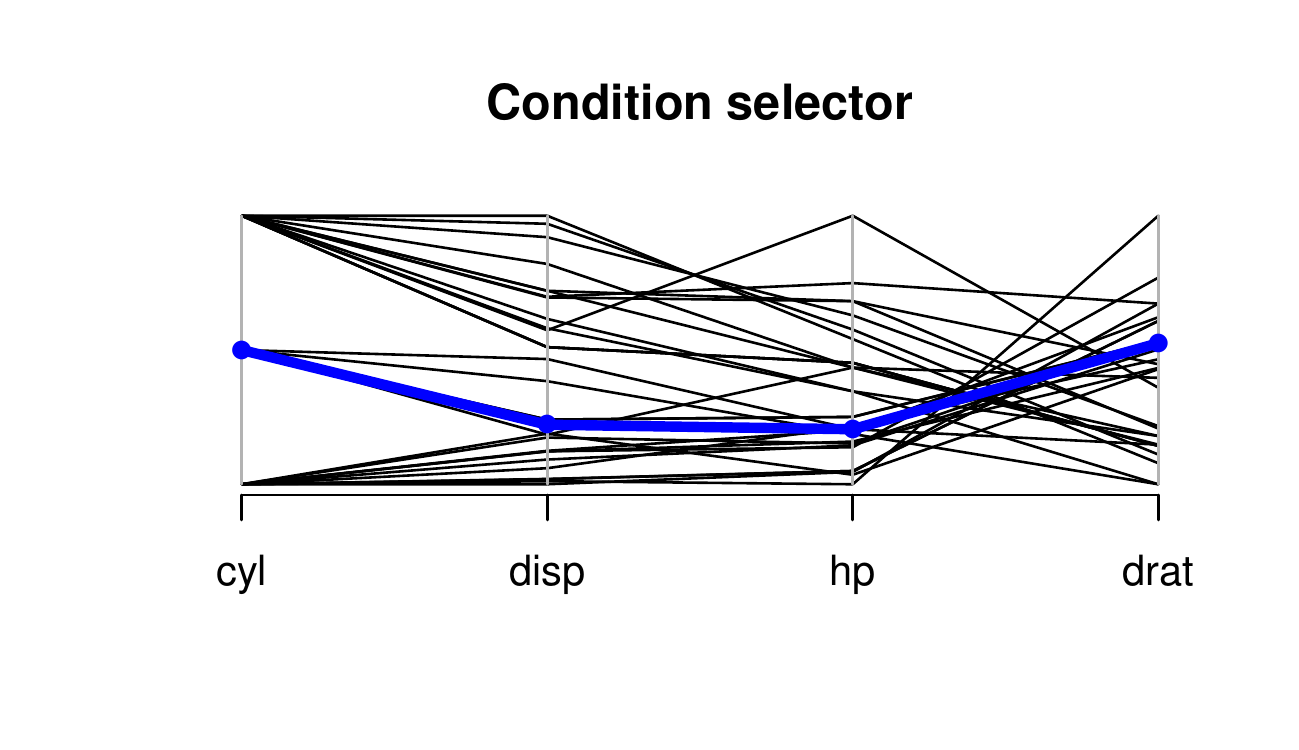}}
  \caption{Full scatterplot and parallel coordinates condition selector plots.}
  \label{fig:otherxc}
\end{figure}

We give three ways to graphically choose a section.
\begin{itemize}
  \item Univariate and bivariate displays: histograms, barplots, scatterplots,
  boxplots, and spineplots (see Figures~\ref{fig:xcuniplots}~and
  ~\ref{fig:xcbiplots}). These are the default condition selectors, because they
  are easy to use and provide some way to avoid the worst extrapolations. The
  scatterplot condition selector is changed to a 2-D histogram when there are
  more than 2,000 observations and the \pkg{gplots} \citep{gplots} package  is
  installed (see Figures~\ref{fig:powerplant1}~and~\ref{fig:powerplant2}).
  \item Full scatterplot matrix (see Figure~\ref{fig:otherxc}). This provides
  every possible bivariate view of the data. It provides more views in which to
  detect extrapolation, but can be very cumbersome. Factors are coerced to
  integers for visualization.
  \item Parallel coordinates plot (see Figure~\ref{fig:otherxc}). Relatively
  neat and easy to understand, but spotting extrapolations is not
  straightforward. Factors are coerced to integers for visualization.
\end{itemize}
\subsection{Ordering condition selector plots}
\label{sec:Corder}
For each approach to graphically choosing sections, we have the choice to order
or group the conditioning predictors to provide for effective exploration of the
predictor space. For bivariate displays, we arrange the conditioning predictors
in pairs using a greedy algorithm that aims to reduce the chance of unwitting
extrapolation (when compared to using univariate views).
\par
For example, in comparing a scatterplot option to the alternative of two
histograms, we consider the ratio of the area of the convex hull of the data in
the bivariate view to the area of the bounding rectangle. If this ratio is near
1, the scatterplot does not offer any extra help over two histograms in avoiding
extrapolations. As this ratio gets smaller, it becomes more important to take
account of the bivariate relationship. In the same manner, when comparing a
spineplot option to the alternative of two barplots, we consider the ratio of
the number of observed factor combinations  to the total possible number of
factor combinations.
\par
This is a simple approach, intended only to give a default ordering --
independent of the model or response -- for users who have not supplied an
ordering. We anticipate that individual users will provide their own ordering,
based on variable importance measures or pre-existing knowledge of the
variables. For example, with a motor insurance pricing model, a salesperson may
wish to have quick access to the effect of voluntary excess or engine size
(inputs which may be changed) in negotiating a policy, but may not have any need
to understand the effect of age or claims history (inputs which cannot be
changed).
\subsection{Visualizing sections}
\label{sec:sections}
\begin{figure}
  \centering
  \vspace{3mm}
  \raisebox{-5mm}{\rotatebox{90}{response}}
  \begin{tabular}{r | c | c | c | c | c}
    \raisebox{+6.5mm}{cont}
    & \includegraphics[width = 18mm]{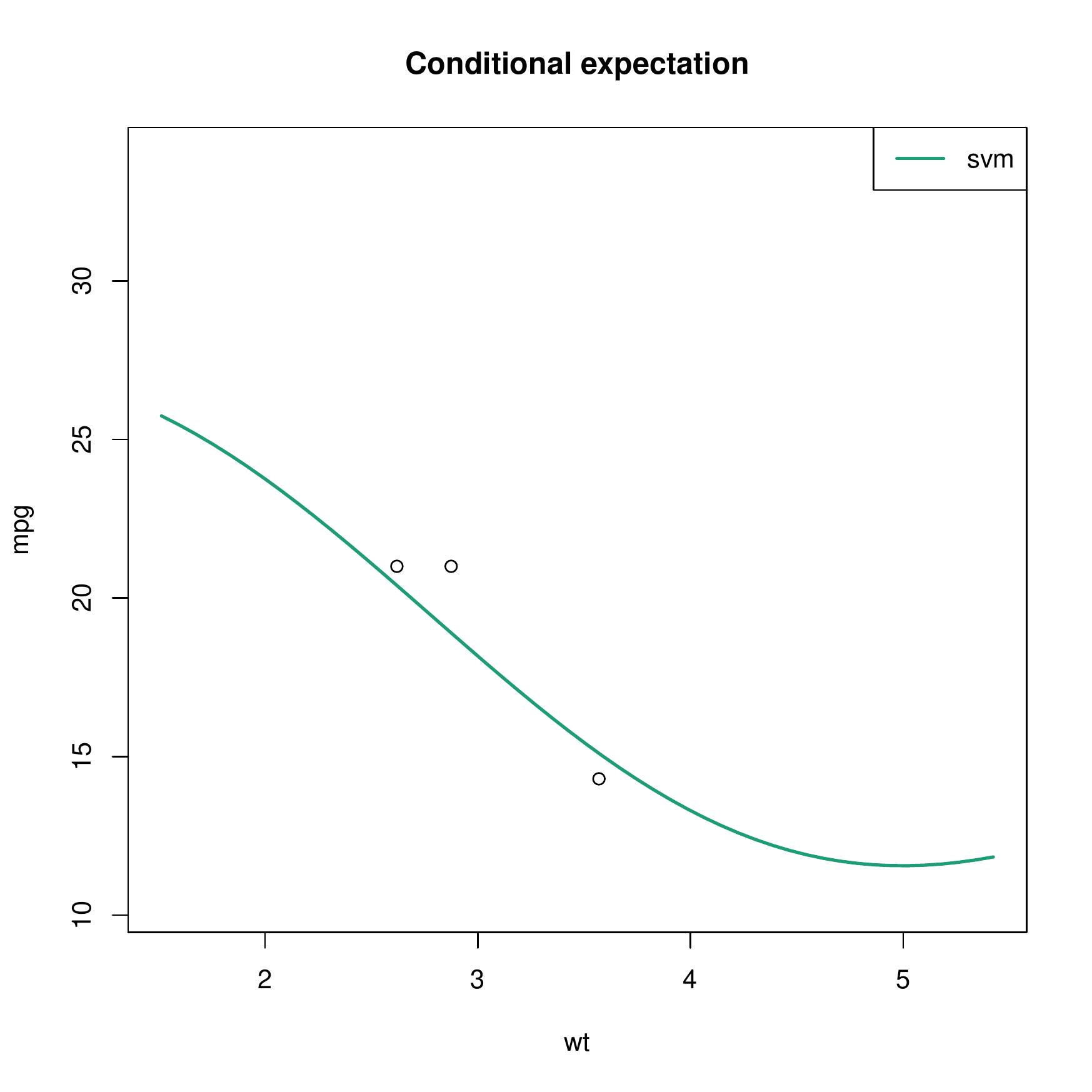}
    & \includegraphics[width = 18mm]{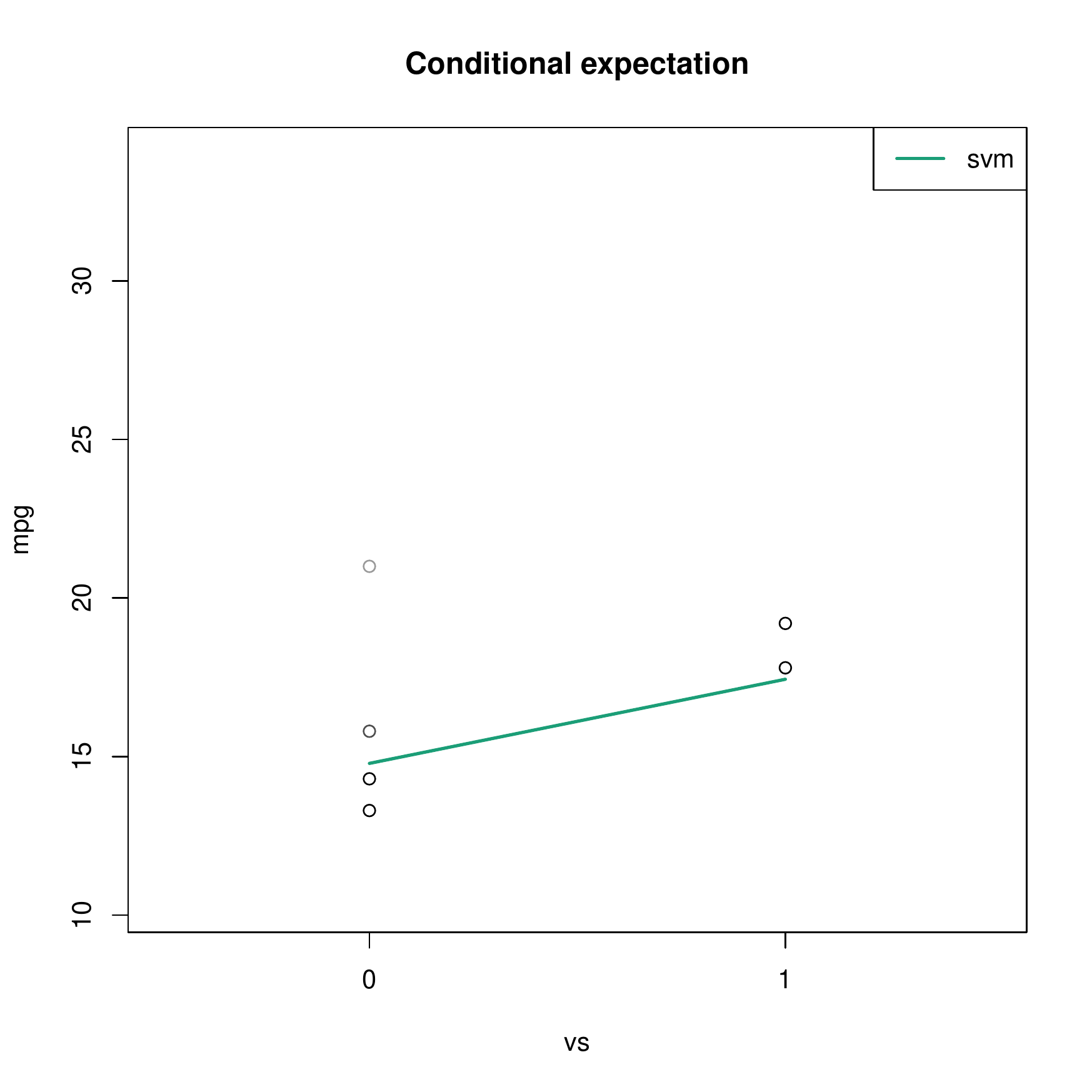}
    & \includegraphics[width = 18mm]{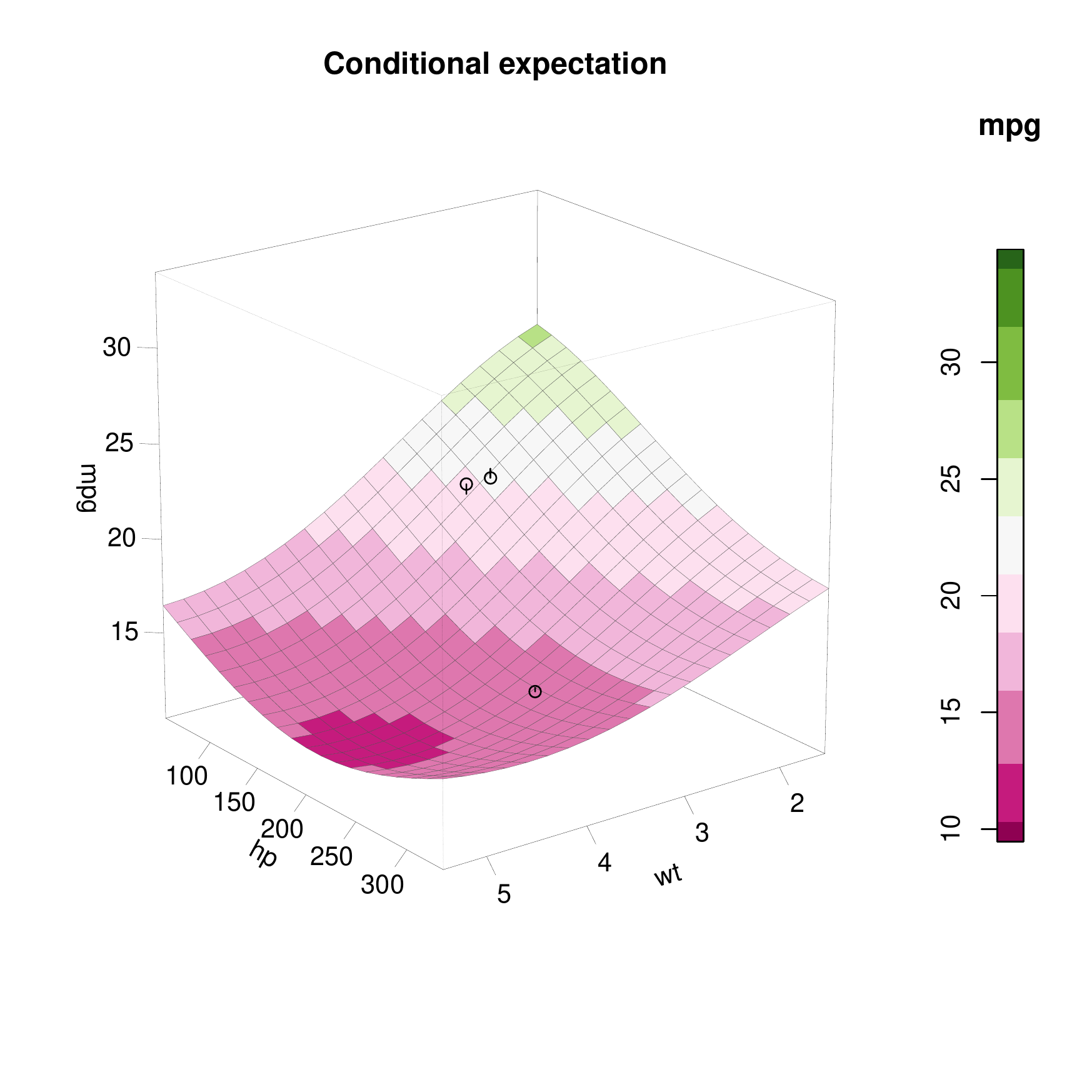}
    & \includegraphics[width = 18mm]{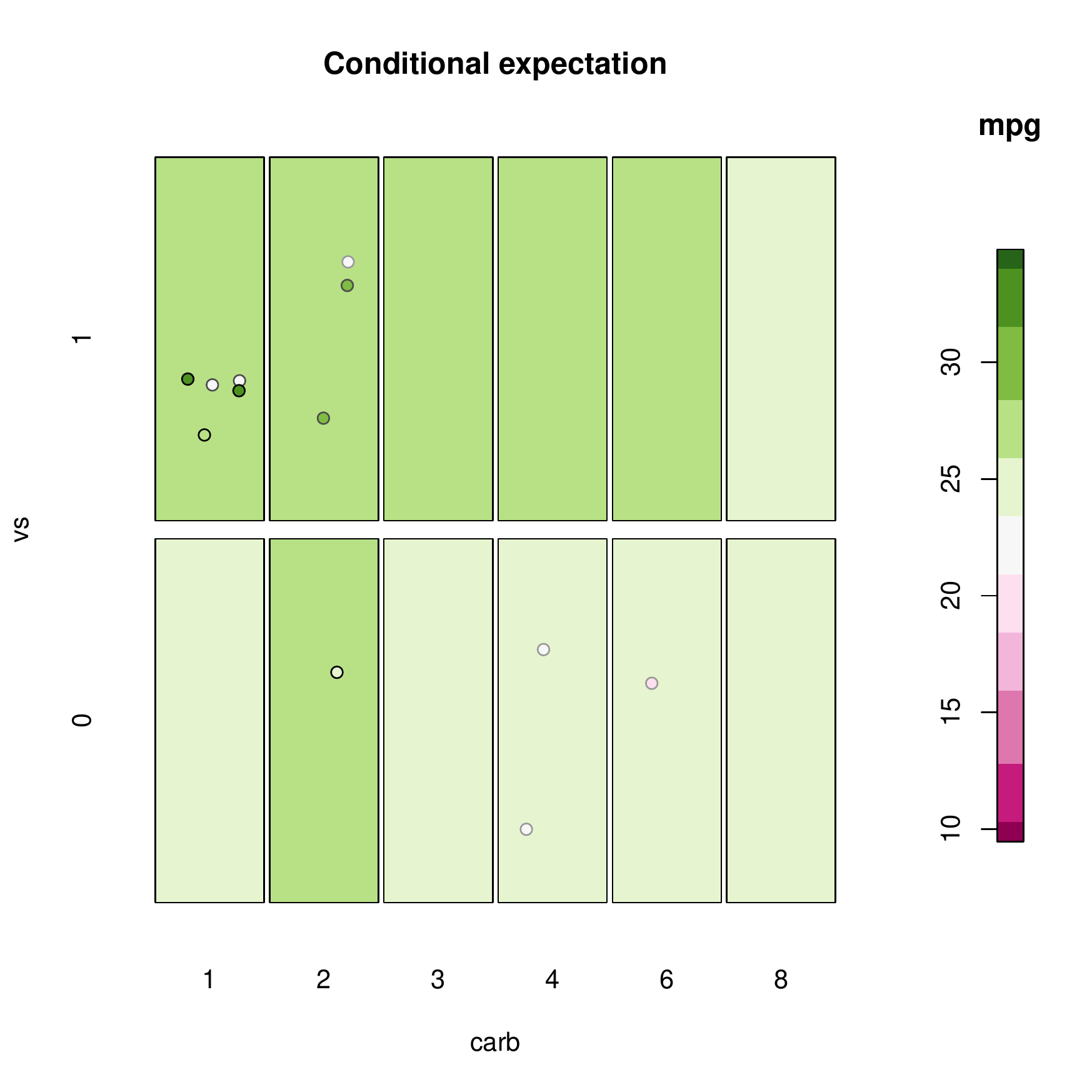}
    & \includegraphics[width = 18mm]{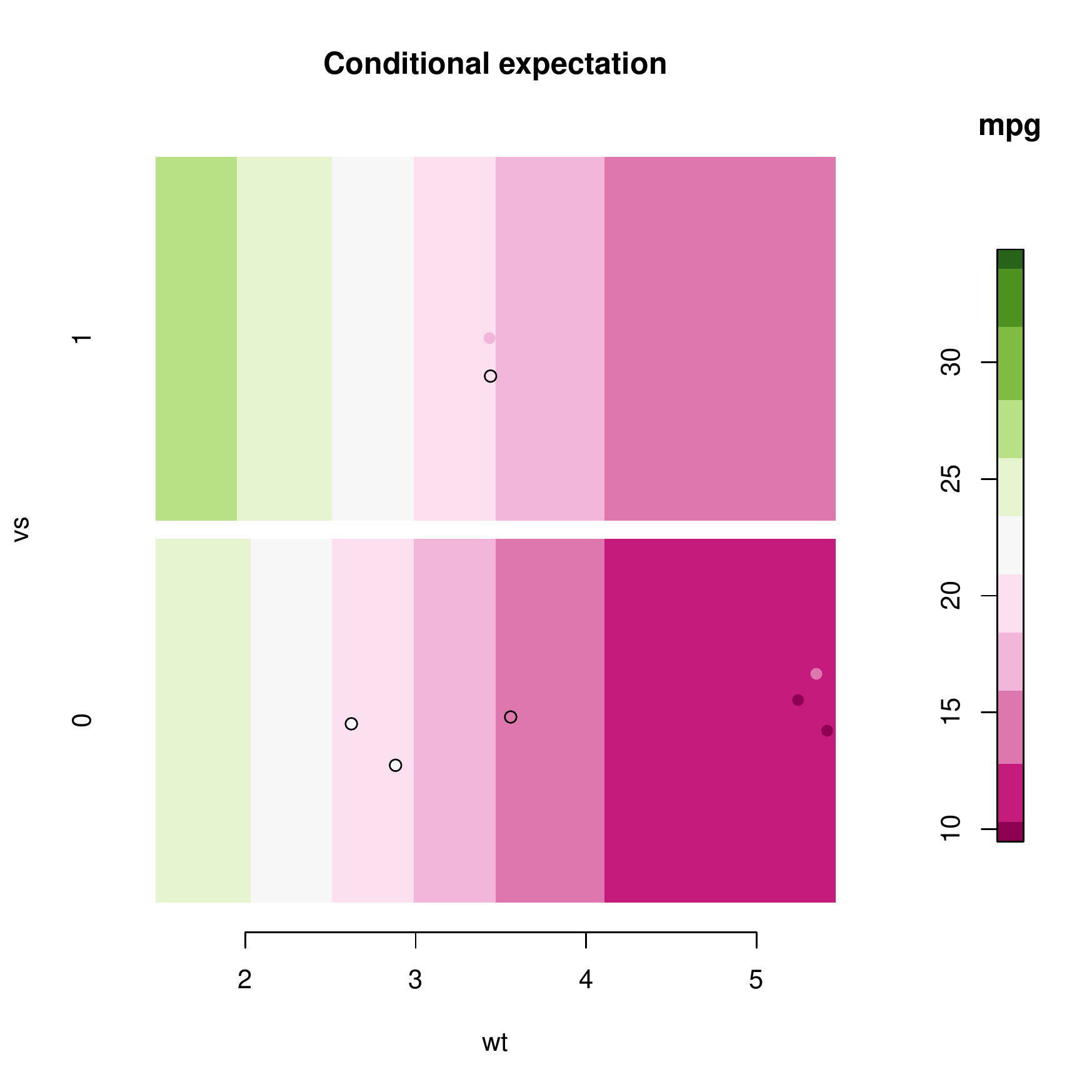}
    \\ \hline
    \raisebox{+6mm}{cat}
    &
    &
    & \includegraphics[width = 18mm]{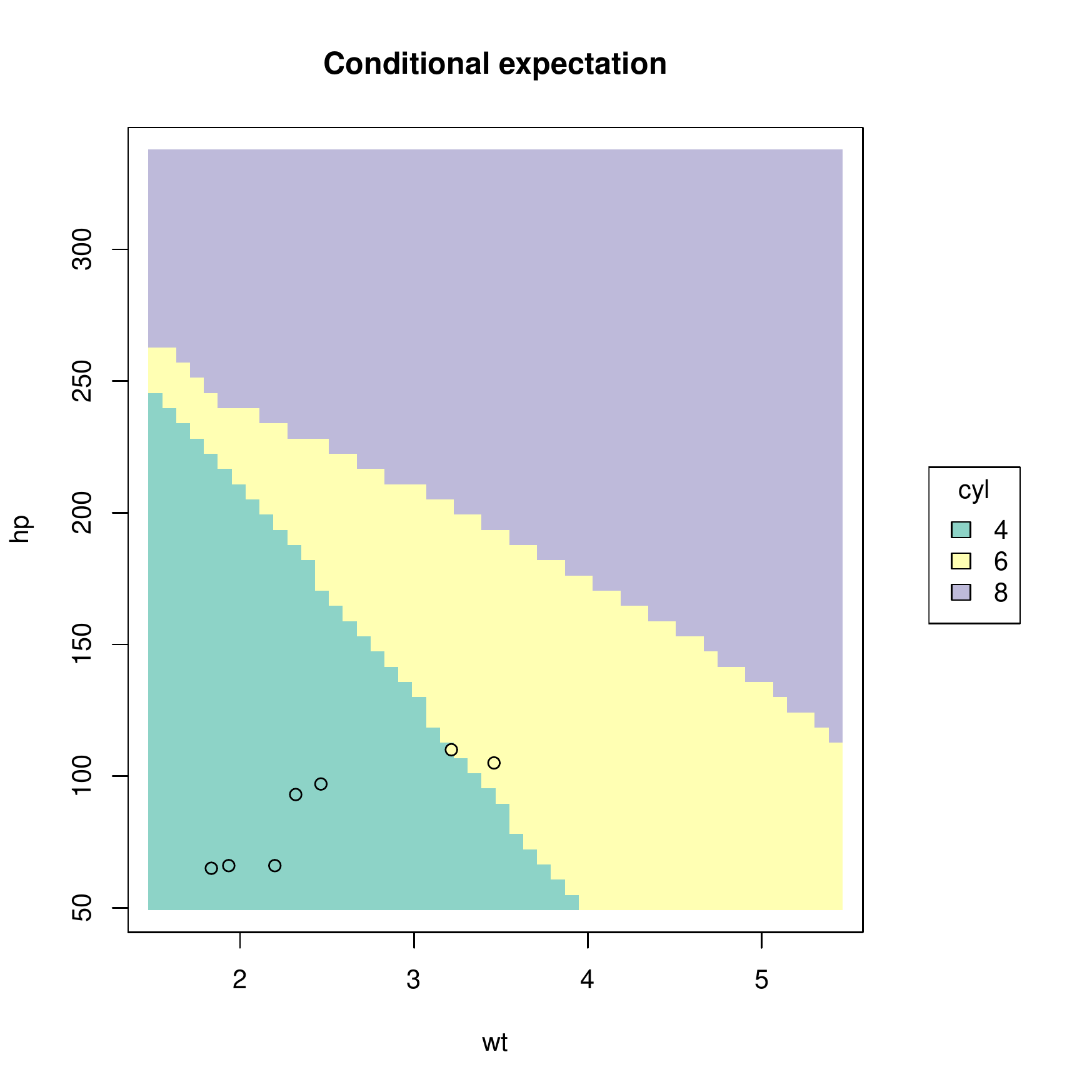}
    & \includegraphics[width = 18mm]{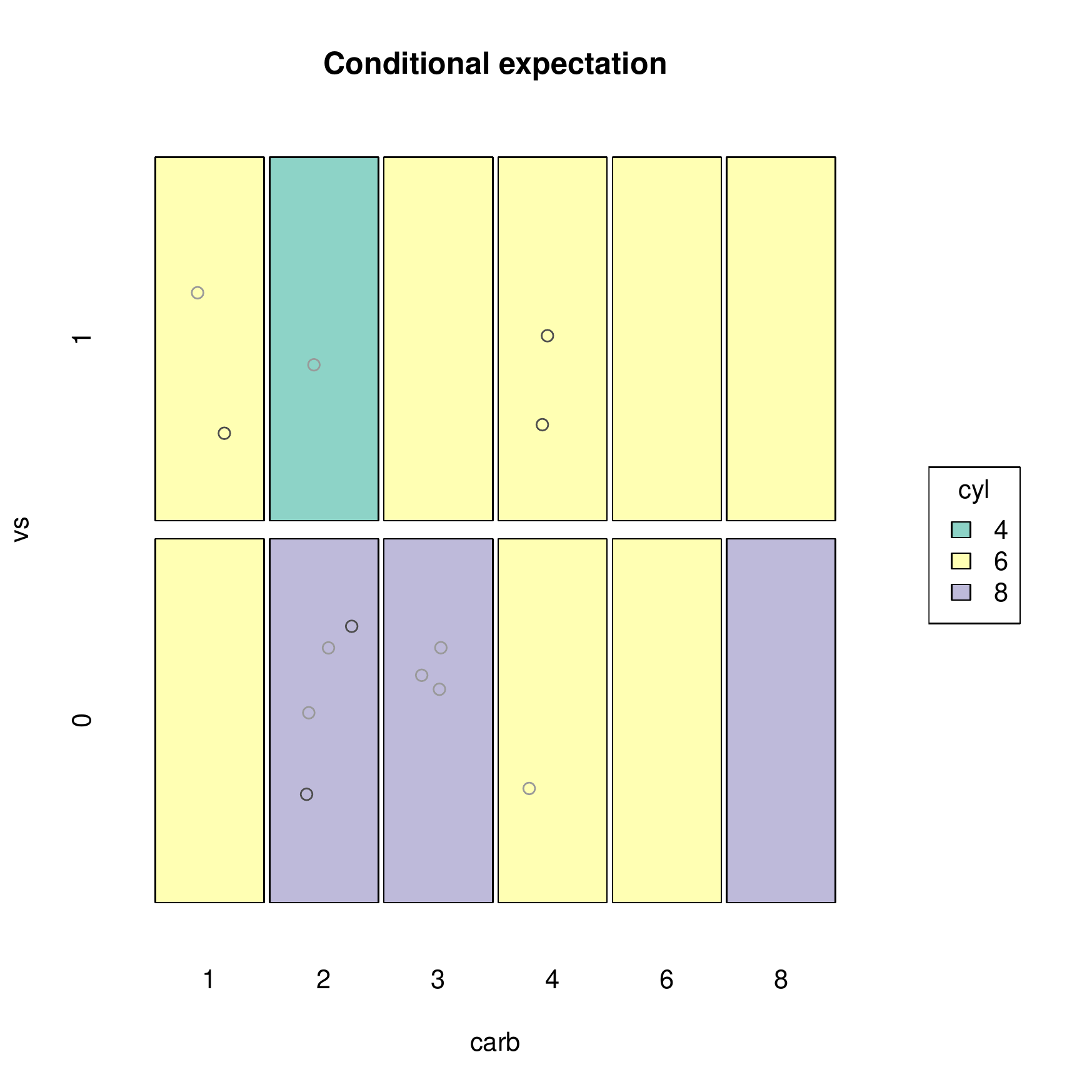}
    & \includegraphics[width = 18mm]{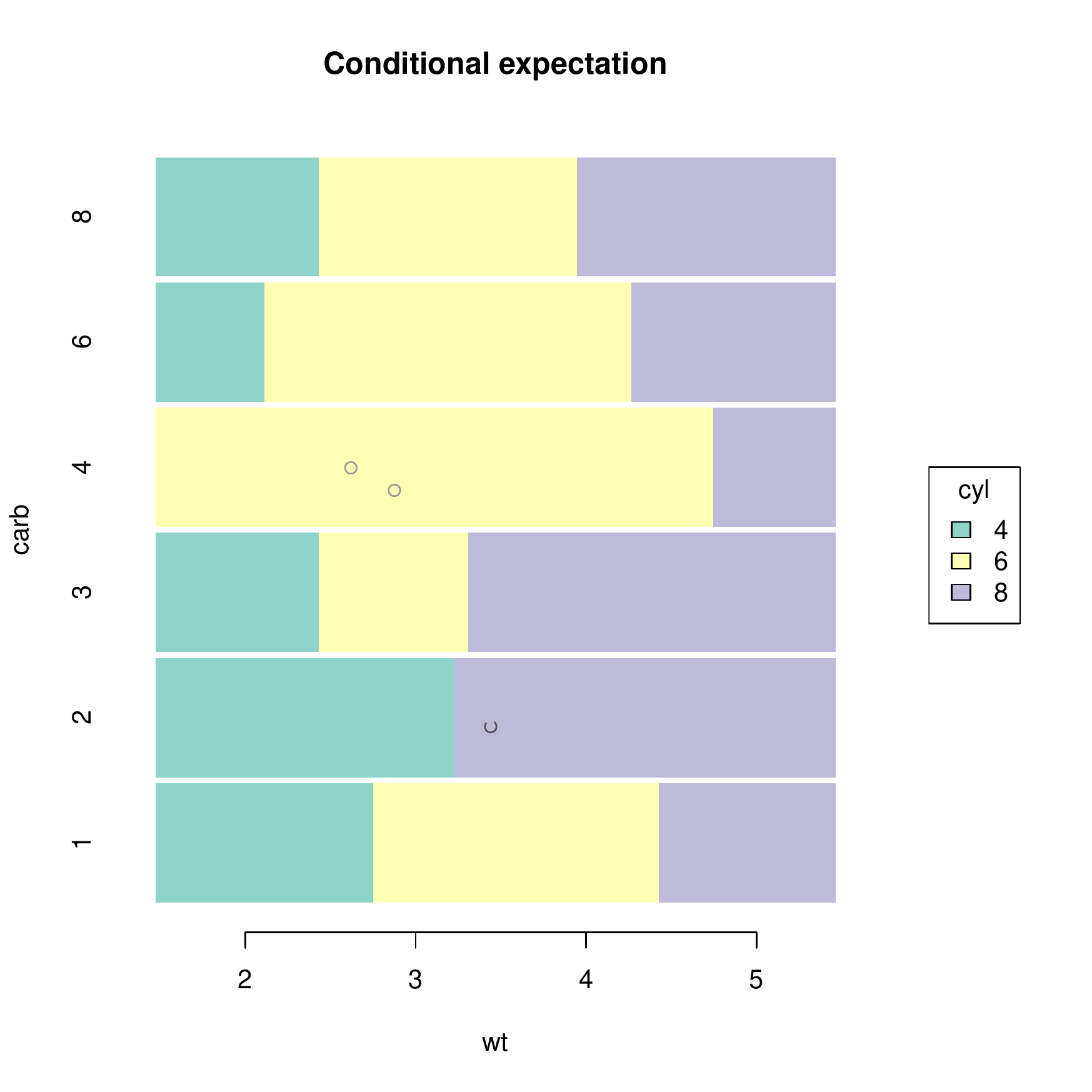}
    \\ \hline \hspace{5mm}
    & 1 cont
    & 1 cat
    & 2 cont
    & 2 cat
    & 1 cont 1 cat
  \end{tabular}
  \raisebox{-5mm}{section predictors}\\
  \caption{Visualizing sections through the fitted model.}
  \label{fig:sections}
\end{figure}

We limit ourselves to conditioning on all but one or two predictors, so we
visualize the intersection of sections and fitted models as either curves or
surfaces (categorical responses are shown as colours, or converted to integers
for visualization). The section visualizations used are summarized in
\mbox{Figure~\ref{fig:sections}}.
The visualizations are chosen according to the type of response, and the type of
predictors we are interested in investigating. Two of the cells in
\mbox{Figure~\ref{fig:sections}} are left blank because with binary logistic
regression, the probability (continuous) of class membership might actually be a
more suitable response to visualize rather than the expected class. If we are
only interested in expected class membership, then it is better to visualize a
section along two predictors. In \mbox{Section~\ref{sec:fev}}, we show an
example of a section with a continuous response and a categorical predictor of
interest.
\par
After we visualize fitted models on the section, we must choose which observed
data to display. We choose observed data according to their distance from the
section.
\par
As in Section~\ref{sec:choosingsections}, we consider both the section and
observed data as points in the space of the conditioning predictors. As a
result, we can take a dissimilarity measure defined between two points as the
dissimilarity between the section and an observed data point
\begin{align*}
  d(\predictors_i, \predictors') = \lpnorm{\predictors_i - \predictors' }{p} +
  \lambda M(\predictors_i, \predictors' )
\end{align*}
where $\predictors_i$ represents the $i$th observation on the conditioning
predictors, and $\predictors'$ gives the current section in the conditioning
predictors. $\lpnorm{\predictors_i - \predictors' }{p}$ is the generalized
Minkowski distance on the numeric elements, $M(\predictors_i, \predictors')$ is
the number of mismatches on the categorical elements, and $\lambda$ is a scaling
constant (argument \code{lambda} to \code{ceplot}). This distance is calculated
after standardizing the continuous elements of the conditioning predictors to
have zero mean and unit variance.
\par
We then have a function which assigns colours to observations based on their
distance from the section:
\begin{align*}
  K(\predictors_i, \predictors', \sigma) = \left\{
  \begin{array}{ll}
  \text{black}
  & \hspace{5.25mm}\mbox{if } 0\leq d(\predictors_i, \predictors') < 0.3\sigma\\
  \text{dark grey}
  & \mbox{if } 0.3\sigma\leq d(\predictors_i, \predictors') < 0.7\sigma \\
  \text{light grey}
  &\mbox{if } 0.7\sigma\leq d(\predictors_i, \predictors') < \sigma \\
  \text{do not plot}
  &\hspace{5mm}\mbox{if } \sigma\leq d(\predictors_i, \predictors')
  \end{array}
  \right.
\end{align*}
where $\sigma$ is a threshold parameter given by the user (argument
\code{threshold} to \code{ceplot}). Letting $\sigma$ equal 0 gives exact
conditioning, that is, we only plot points on the section. Increasing $\sigma$
gives more approximate conditioning. Setting $\lambda > \sigma$ means that only
observations which match the section on categorical predictors may be considered
to be \emph{near} the section. The correct way to achieve this with
\code{ceplot} is to leave \code{threshold = NULL}, which requires all factor
levels to match but is more computationally efficient.
\par
\begin{figure}
  \vspace{8mm}
  \centering
  \hspace{-5mm}
  \begin{subfigure}[b]{0.43\textwidth}
  \includegraphics[width = \textwidth, clip = TRUE, trim = {0 0 0 9mm}]
    {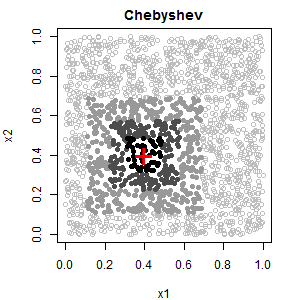}
  \caption{}
  \end{subfigure}
  \hspace{5mm}
  \begin{subfigure}[b]{0.43\textwidth}
  \includegraphics[width = \textwidth, clip = TRUE, trim = {0 0 0 9mm}]
    {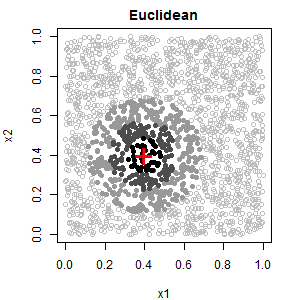}
  \caption{}
  \end{subfigure}
  \caption{Choosing data to display using distance between section and
  observations. Condition on x1 and x2 by taking a section defined by x1 = x2 =
  0.4. There are no points exactly at these coordinates, so we take points
  nearby. (a) Maximum norm distance. (b) Euclidean distance.}
  \label{fig:distance}
\end{figure}

The \pkg{condvis} package currently implements two special cases of the
Minkowski distance, namely maximum norm and Euclidean distance. The maximum norm
is the limit of the Minkowski distance as $p$ tends to infinity
\begin{align*}
  \lpnorm{\predictors_i - \predictors' }{\infty} = \max_{j} \lvert \predictors_
    {ij} - \predictors'_j \rvert
\end{align*}
The maximum norm results in conditioning very much like that of trellis
\citep{Becker96trellis}, where intervals on conditioning predictors are used to
choose observations to plot on a single panel. Euclidean distance is the
Minkowski distance with $p$ equal to 2. See Figure~\ref{fig:distance} for an
example of both distance measures with two conditioning predictors.
\subsection{Interactive graphics platform}
The interactive graphics in \pkg{condvis} have been implemented using base
\proglang{R} graphics. Extensive use is made of \code{getGraphicsEvent} and
related functions in \pkg{grDevices} \citep{R}, which were introduced in
\proglang{R 2.1.0}, released in April 2005. The decision to develop
\pkg{condvis} on this platform was motivated chiefly by the desire to keep
software dependencies to a minimum. The main non-standard dependency of
\pkg{condvis} is the XQuartz (\href{http://www.xquartz.org/} {www.xquartz.org})
device on the Mac OS.
\par
The secondary benefit of developing interactive graphics in \proglang{R} is the
huge wealth of static graphics which are already available in both the standard
\proglang{R} distribution and its extension packages. If we can interact with
the \proglang{R} graphics device, we do not need to port these plots to another
language/platform in order to interact with them. In this package alone, we take
advantage of the following graphics: histogram, barplot, scatterplot, boxplot,
spineplot, 2-D histogram, 3-D perspective mesh, and parallel coordinates plot.
\par
The Shiny web applications implemented in \pkg{condvis} provide an alternative
to the default implementations, and allow the graphics to be deployed to a web
page for others to use. On Mac OSX, Shiny becomes the default for \code{ceplot}
in the absence of XQuartz. See Section~\ref{relatedinR} for further discussion
of interactive graphics.
\newpage
\section{Using the package}
\label{sec:using}
\subsection{Graphic type and layout}
\label{sec:layout}
There are three layout options for the interactive graphics created by
\code{ceplot}:
\begin{itemize}
  \item The default option is to place the section and condition selector plots
  on one device, with the section being the largest graphic on the left, and the
  condition selector plots arranged in columns on the right.
  \item The separate\footnote{Special care should be taken when using this
  option with R version $<$ 3.2.2, as closing the non-interactive device (the
  one showing the section) can cause a crash (see
  \href{https://bugs.r-project.org/bugzilla3/show_bug.cgi?id=16438}{Bug 16438 at
  \mbox{bugs.r-project.org}})} option is set using \code{type = "separate"} when
  calling \code{ceplot}. This places the section on one device, and the
  condition selector plots on a second device. This option gives the flexibility
  to use the scatterplot matrix and parallel coordinates plots for condition
  selectors.
  \item The Shiny option is set using \code{type = "shiny"} when calling
  \code{ceplot}. This gives an arrangement similar to the default above, but is
  implemented as a web application. This option allows some extra interactivity,
  such as changing the distance function type and threshold value, and the
  ability to deploy your interactive graphic to the internet for sharing.
\end{itemize}
\subsection{Interacting with the graphics}
The main user interaction with \pkg{condvis} is through clicking on condition
selector plots in order to select sections to visualize. When the section being
visualized is a three-dimensional perspective mesh, it may be rotated using the
arrow keys. In the default plot, the perspective mesh may also be rotated by
clicking and dragging with the mouse. In Shiny, the rotation may instead be
adjusted using sliders. When using the default or separate plots, a snapshot of
the current visualizations may be taken by pressing the `s' key. A pdf snapshot
is then saved to the working directory. In Shiny, the button for this is
labelled ``Download snapshot (pdf)''.
\subsection{Examples}
The following examples use the FEV data again, as well as the powerplant data
from \citet{Tufeckipowerplant}, and the wine data from \citet{winedata}.
\subsubsection*{FEV data}
\begin{figure}[ht!]
  \vspace{8mm}
  \centering
  \includegraphics[width = 0.65\textwidth, clip = TRUE, trim = {0 0 0 14mm}]
    {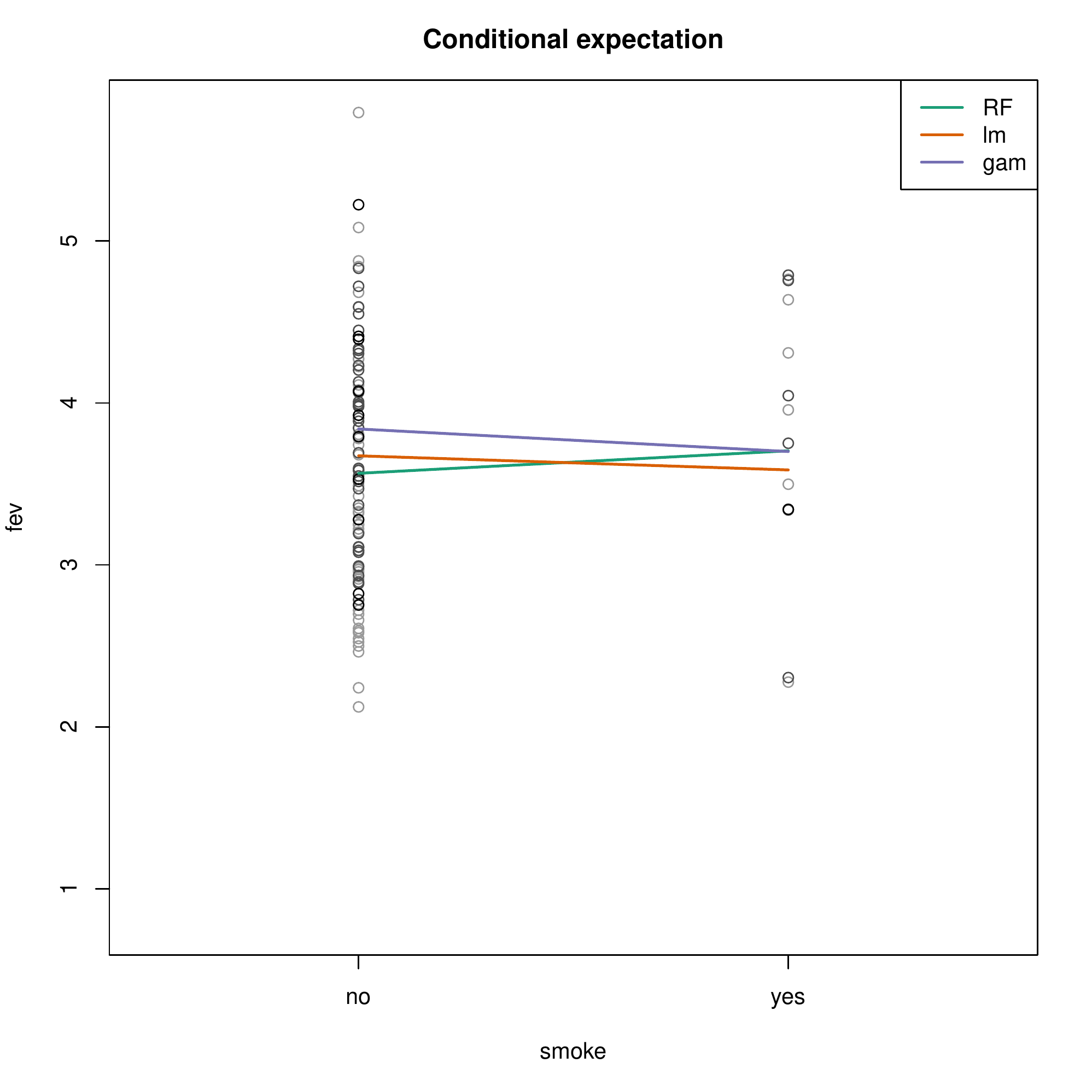}
  \raisebox{+25mm}{\includegraphics[width = 0.2\textwidth]{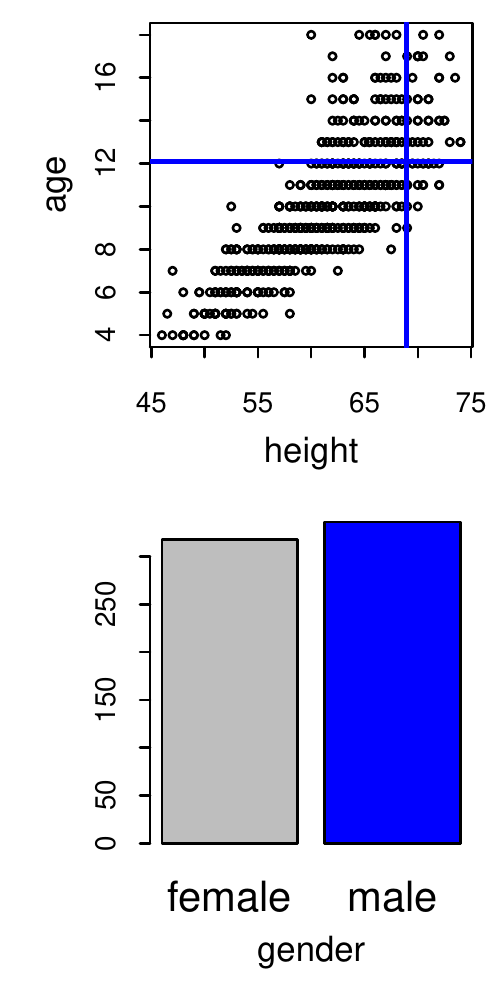}}
  \caption{Snapshot comparing the modelled effect of smoking on FEV from three
  different models.}
  \label{fig:fev1}
  \vspace{5mm}
\end{figure}

We use the FEV dataset from Section~\ref{sec:fev} for another example here. This
time, we are not addressing any new question, rather we are just exploring a few
different fitted models. This example shows the default plot arrangement on a
single device, and the comparison of several models on a 2-D section.
\begin{Code}
R> library("randomForest")
R> library("mgcv")
R> m3 <- list(
+   RF = randomForest(fev ~ ., data = fev),
+   lm = lm(fev ~ ., data = fev),
+   gam = mgcv::gam(fev ~ smoke + gender + s(age) + s(height),
+     data = fev))
R> ceplot(data = fev, model = m3, sectionvars = "smoke")
\end{Code}
See Figure~\ref{fig:fev1} for a snapshot of the interactive graphic resulting
from this call to \code{ceplot}.
\subsubsection*{Powerplant data}
The powerplant data were collected for the purpose of predicting full load
electrical power output of a powerplant. The underlying processes are well
understood, but require the difficult computation of differential equations in
order to model them. \citet{Tufeckipowerplant} suggests machine learning
techniques as an alternative and proceeds to fit several different `black-box'
models. One interesting aspect of this article is that, in presenting these
complex models to an arguably non-statistical audience, there are no graphics
produced to visualize a predictor effect.
\par
\begin{figure}
  \centering
  \includegraphics[width = 0.65\textwidth, clip = TRUE, trim = {0 0 0 14mm}]
    {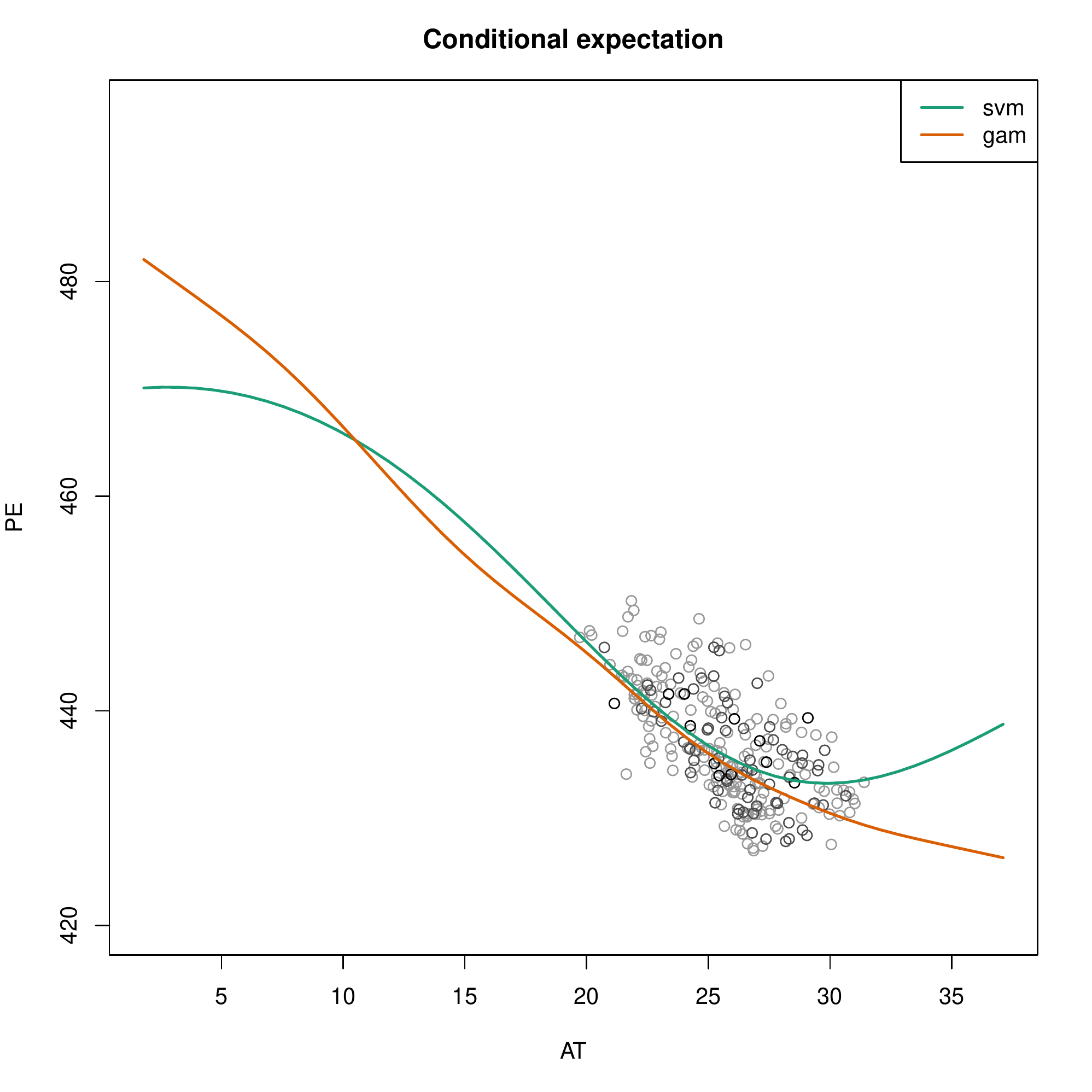}
  \raisebox{+25mm}{\includegraphics[width = 0.2\textwidth]
    {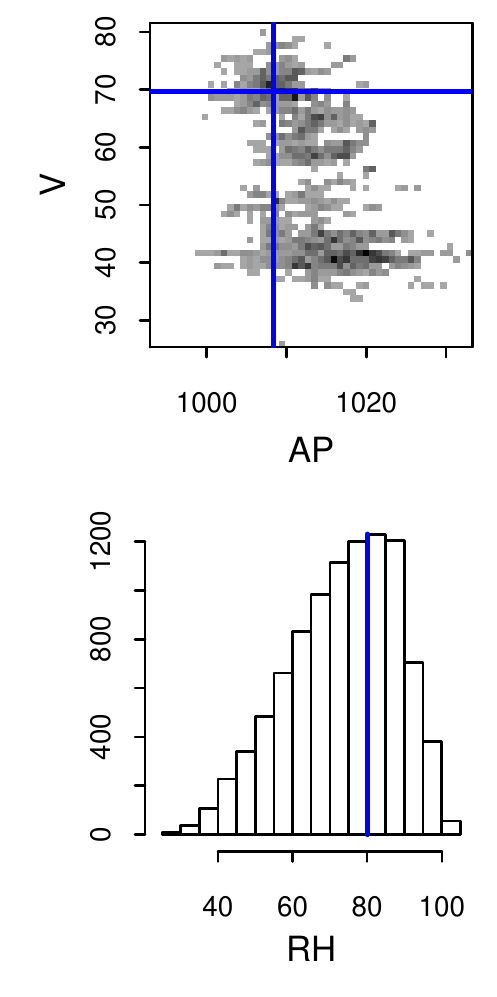}}
  \caption{Snapshot showing the modelled effect of \code{AT} given some fixed
  values of \code{V}, \code{AP} and \code{RH}, for two different models. The two
  models look quite different, except in the region where observed data are
  nearby.}
  \label{fig:powerplant1}
\end{figure}
\begin{figure}
  \centering
  \includegraphics[width = 0.65\textwidth, clip = TRUE, trim = {0 0 0 15mm}]
    {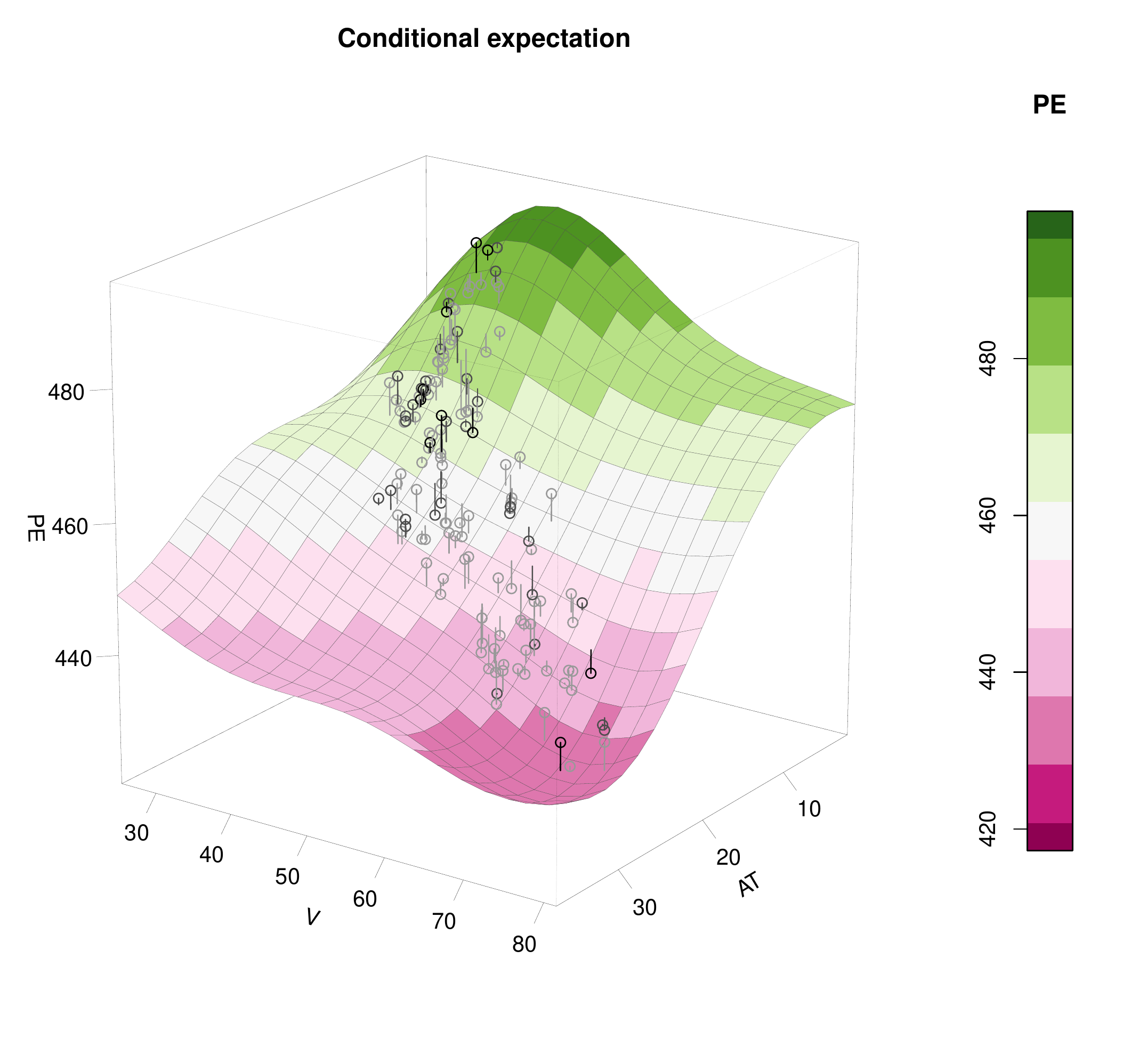}
  \raisebox{+50mm}{\includegraphics[width = 0.2\textwidth]
    {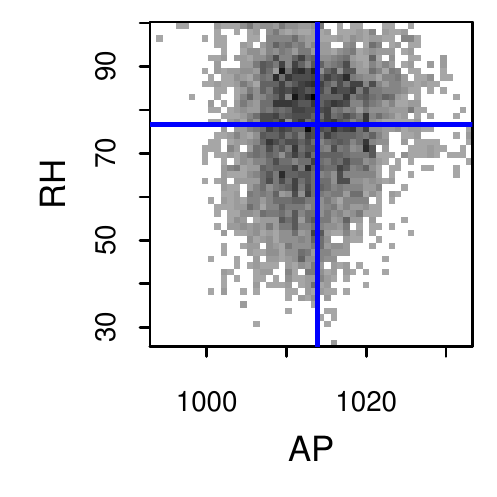}}
  \caption{Snapshot showing the modelled effect of \code{V} and \code{AT} given
  some fixed values of \code{RH} and \code{AP}. Note the condition selector is a
  two-dimensional histogram rather than a scatterplot.}
  \label{fig:powerplant2}
\end{figure}

We fit two models to the powerplant data -- a support vector machine with a
radial kernel, and an additive model with smoothing splines for each predictor.
\newpage
\begin{Code}
R> data("powerplant")
R> library("e1071")
R> library("mgcv")
R> m4 <- list(
+    svm = svm(PE ~ ., data = powerplant),
+    gam = gam(PE ~ s(AT) + s(V) + s(AP) + s(RH),
+      data = powerplant))
\end{Code}
The first call to \code{ceplot} shows how the support vector machine fits the
data, showing how some of the modelled effect curvature occurs in regions of the
data space with no observed data.
\begin{Code}
R> ceplot(data = powerplant, model = m4["svm"], sectionvars = "AT",
+    type = "separate")
\end{Code}
The next call to \code{ceplot} compares the support vector machine to the
additive model (Figure~\ref{fig:powerplant1}, or Shiny
\href{https://markajoc.github.io/condvis/example-powerplant.html}{demo}),
showing how both models fit the data in a similar way in regions of the data
space near observed data.
\begin{Code}
R> ceplot(data = powerplant, model = m4, sectionvars = "AT",
+    type = "separate", threshold = 0.5)
\end{Code}
The final call to \code{ceplot} demonstrates a 3-D section.
\begin{Code}
R> ceplot(data = powerplant, model = m4["svm"],
+    sectionvars = c("AT", "V"), type = "separate", view3d = TRUE,
+    threshold = 0.2)
\end{Code}
See Figure~\ref{fig:powerplant2} for a snapshot of this example. There is a
video demonstration of this example
\href{https://www.youtube.com/watch?v=rKFq7xwgdX0&feature=youtu.be&t=846}{here}.
\subsubsection*{Wine data}
The wine data are the results of a chemical analysis of wines grown in the same
region in Italy but derived from three different cultivars. The basic task is to
produce a classifier which can take the chemical measurements and identify the
correct cultivar. For illustration purposes, we fit a \code{randomForest}
\citep{randomForest-package} classifier on six of the predictors. (Shiny
application \href{https://markajoc.github.io/condvis/example-wine.html}{demo}.)
\begin{Code}
R> library("randomForest")
R> data("wine")
R> wine$Class <- as.factor(wine$Class)
R> m5 <- randomForest(Class ~ Alcohol + Malic + Ash + Magnesium +
+    Phenols + Flavanoids, data = wine)
\end{Code}
It is difficult to comprehend how a classifier assigns regions to different
classes in six dimensions, but visualizing this on a two-dimensional section is
straightforward. We first take sections along the predictors \code{Alcohol} and
\code{Phenols}, using the option to create a Shiny web application.
\begin{Code}
R> ceplot(data = wine, model = m5, sectionvars = c("Alcohol", "Phenols"),
+    type = "shiny")
\end{Code}
Next, we visualize sections on the same predictors, but use the parallel
coordinates condition selector.
\begin{Code}
R> ceplot(data = wine, model = m5, sectionvars = c("Alcohol", "Phenols"),
+    type = "separate", selectortype = "pcp", threshold = 2)
\end{Code}
\begin{figure}
  \centering
  \includegraphics[width = 0.95\textwidth]{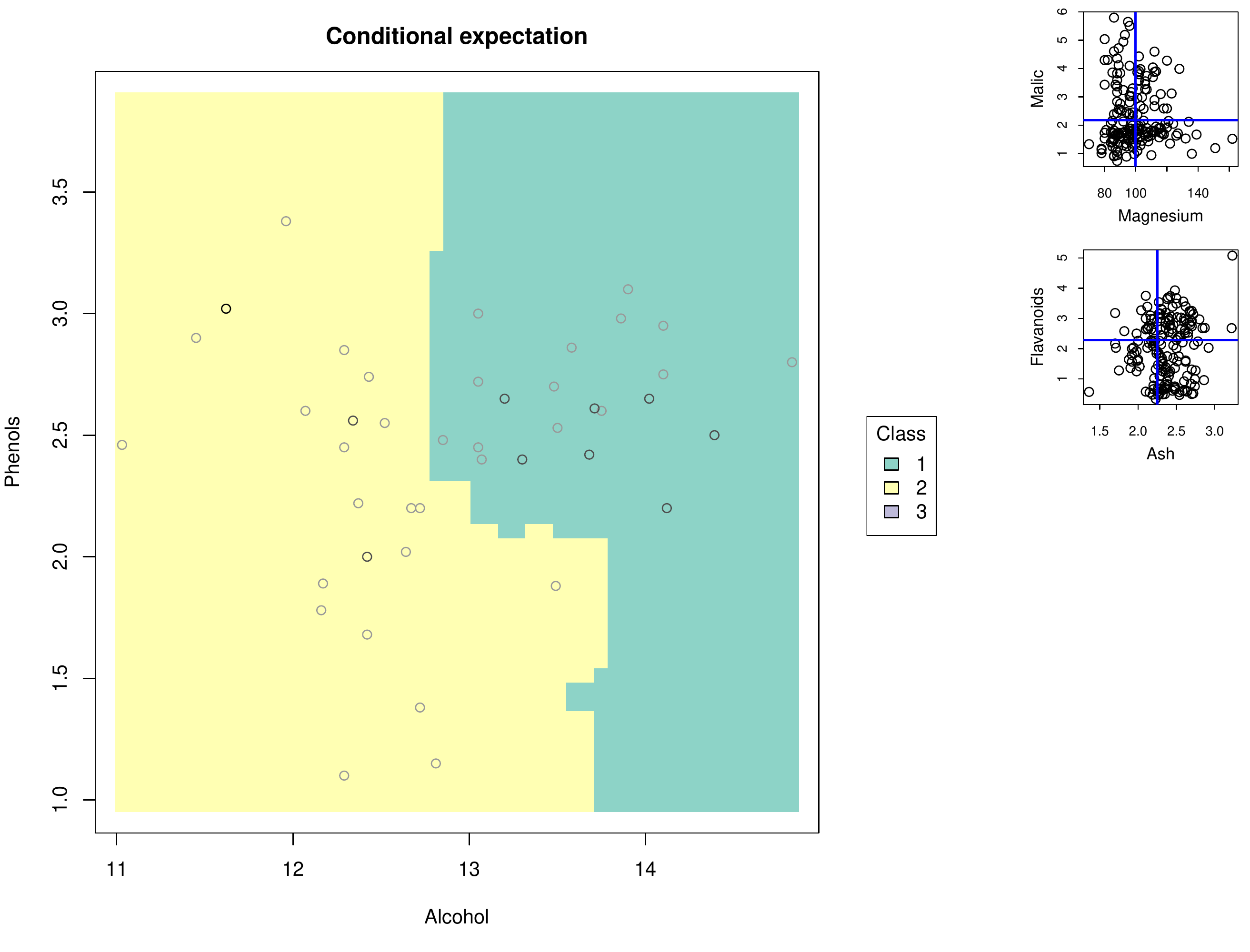}
  \caption{Snapshot showing a section through a random forest classifier in six
  dimensions. Shiny application
  \href{https://markajoc.github.io/condvis/example-wine.html}{here}.}
  \label{fig:wine}
\end{figure}
\section{Related work}
\label{sec:related}
\subsection{Related work in the literature}
Trellis graphics \citep{Becker96trellis} are separate plots visualizing subsets
of the data. Conditioning on continuous variables is achieved by the
intersection of intervals on each variable. \citet{Nason04CARTscans} apply the
trellis technique to the conditional visualization of fitted models, calling the
resulting plots CARTscans. Embedded plots \citep{grolemund15embedded} are a
collection of graphs organized in a larger graphic, that can display more
complex relationships than typically possible. \citet{Furnas94prosectionviews}
discuss the use of sections and projections in exploring high-dimensional space.
\par
Partial residual plots \citep{LarsenMcCleary72, ezekiel1924curvilinear} produce
a single plot showing the effect of one or two predictors, conditional on the
remaining predictors and a fitted additive model. \citet{cook93partialresiduals,
cook1996curvature} notes that the model used in the construction of a partial
residual plot must be reasonably well specified for the conditioning predictors
in order to have a valid plot. Partial residual plots are also known as
component + residual plots.
\par
ICE plots \citep{ICEboxpaper2014} show sections through a fitted model at
observed data points. The main aim is to allow interpretation of `black-box'
model effects, and give an impression of extrapolation behaviour. Partial
dependence plots \citep{friedman2001} predate ICE plots, but can be interpreted
as the average of the sections shown in an ICE plot.
\par
Trellis offers a simple, interpretable approach, but it is difficult to apply
it to fitted models, especially with continuous predictors. Partial residual
plots are limited by their dependence on a fitted additive model. ICE plots can
visualize more complicated non-additive models, but run into problems of
overplotting and do not explicitly show the observed data.
\subsection{Related work in R}
\label{relatedinR}
\subsubsection*{Conditional visualization}
The \pkg{lattice} \citep{lattice} package is an implementation of trellis
graphics in \proglang{R}. The \code{coplot} function in \pkg{graphics} \citep{R}
also provides a conditioning plot following the trellis method. The
\pkg{effects} \citep{effects} package produces partial residual plots and
related graphics for additive models. The \pkg{visreg} \citep{visreg} package
produces partial residual plots for additive models as well as extensions to the
non-additive case. The plot method \code{plot.gam} in both \pkg{gam}
\citep{gam-package} and \pkg{mgcv} \citep{mgcv} allows the production of
two-dimensional and three-dimensional partial residual plots for models of class
\code{gam}. The \pkg{ICEbox} \citep{ICEboxpaper2014} package implements ICE
plots. These are all static visualizations.
\subsubsection*{Interactive graphics}
Many approaches to interactive graphics in \proglang{R} port the graphics to
another language for the interactive functionality, for example: \pkg{iplots}
\citep{iplots-package}, \pkg{shiny} \citep{shiny}, \pkg{ggvis}
\citep{ggvis-package}. The main approach in \pkg{condvis} is to interact
directly with graphics produced by \proglang{R} code, displayed on standard
\proglang{R} graphics devices. This is achieved using functions contained in the
\pkg{grDevices} package \citep{R}, which is distributed with the basic
\proglang{R} installation. The use of \pkg{grDevices} for interactive graphics
can also be seen in the Association Navigator of \citet{buja2010assocnav}, and
the \pkg{sudoku} \citep{sudoku-package} package. The supplementary material
contains three \proglang{R} scripts to demonstrate the use of \pkg{grDevices} in
producing basic interactive graphics with \proglang{R}.
\subsection{Contribution}
The \pkg{condvis} package takes the approach of visualizing the model in data
space (see discussion by \citet{wickham15blindfold}). Out of the related methods
mentioned previously, \pkg{condvis} is most similar to CARTscans
\citep{Nason04CARTscans} in the sense of applying conditional visualization to
fitted models. Rather than using trellis graphics for conditioning, we take a
more flexible, interactive approach to conditioning. Instead of producing
multiple plots representing `fat' sections or subregions of the data space, we
produce one section at a time and attempt to show data which are near this
section. This means that fitted models are represented exactly (no need for
integration or averaging of the fitted model), and we also display observed data
to give context to the section.
\section{Summary and outlook}
\label{sec:summary}
\subsection{Summary}
The \pkg{condvis} package allows the user to interactively take 2-D and 3-D
sections in data space and visualize fitted models where they intersect the
section, and observed data if sufficiently near the section according to a
distance measure.
\subsection{Strengths and limitations}
The strength of \pkg{condvis} lies in creating low-dimensional visualizations
of fitted models in high dimensional space. This allows interpretation of
complex `black-box' models, and shows how observed data actually support the
fitted model in data space. The software has been tested and shown to work on
Windows, Mac and Linux, with the following model classes in \proglang{R}:
\code{lm}, \code{glm}, \code{gam}, \code{svm} (\pkg{e1071}), \code{rpart},
\code{randomForest}.
\par
This method of interactively choosing sections is, however, limited to dealing
with no more than, say 20 or 25 conditioning predictors. The condition selectors
would begin to crowd the screen at this point, and exploring the data space with
only the help of bivariate data visualizations would become difficult. This
method also does not suit situations where categorical predictors have more than
4 or 5 levels. The barplots and spineplots in such cases may not represent these
predictors well for the purpose of choosing conditions.
\subsection{Further work}
This work has already opened up some avenues of further research. It becomes
obvious after a short time using \pkg{condvis} that it can sometimes be
difficult to choose interesting sections, or indeed to feel confident that the
data space can be explored effectively when choosing sections by hand. This
naturally suggests an automated approach to producing sections throughout the
data space.
\par
Such an automated approach would likely result in a very large number of
sections, the direct visualization of which would not be humanly possible. One
solution to this may be to apply the `cognostics' approach as discussed in
\citet{tukey1982}. This involves using a computer to rank plots according to
some criteria, and subsequently choosing a subset of plots for examination by a
human user.
\par
Another approach would be to order the sections according to their location in
the data space, and then produce an animation which transitions between the
sections, resulting in a sort of conditional tour of the space (as in a tour of
sections, rather than a tour of projections as in \citet{asimov1985grandtour}).
This feature is currently in development for \pkg{condvis} (see
\code{?condtour}).
\par
So far, we have discussed conditional visualization as a one-way process; choose
a section/condition and then investigate the section. The inverse problem might
also be interesting, for example, what areas of the predictor space have high
values of expected response, or the highest curvature of the fitted model in a
certain direction?
\par
We have taken a very strict approach to conditioning in this discussion. We have
only considered conditions where predictors are set to a single value. If we
consider conditions more flexibly over regions of the predictor space, the
fitted model can no longer be visualized directly. However, it is possible to
use approximate integrals of the fitted model over such regions to make
informative visualizations.
\par
Finally, only some of our section visualizations provide for model comparison.
As it stands, none of our 3-D sections allow model comparison, and so this is
something we would like to include in further versions of \pkg{condvis}.
\bibliography{jss2576}
\appendix
\linespread{1}
\newpage
\section{Supplementary material}
The supplementary material consists of the source code for \pkg{condvis}, a
replication script for the examples in this article, and three further scripts
which demonstrate the kind of interactive graphics described in the paper.
\begin{itemize}
  \item \textit{condvis\_0.3-4.tar.gz}: source code for \pkg{condvis}, version
  0.3-3 available on CRAN.
  \item \textit{replication.R}: \proglang{R} script for replicating examples in
  this article
  \item \textit{interactive-example-1.R}: Example of interacting with
  a single \proglang{R} plot.
  \item \textit{interactive-example-2.R}: Example of interacting with
  a single \proglang{R} plot, and propagating the change to another plot.
  \item \textit{interactive-example-3.R}: Example of interacting with
  multiple \proglang{R} plots, propagating the changes to the other plots as
  required. This currently seems to work only in Windows.
\end{itemize}
\end{document}